\documentclass[preprint,10pt]{sigplanconf}

\usepackage{verbatim}
\usepackage{xspace}

\usepackage{array}
\usepackage{listings}
\usepackage{fancybox}
\usepackage{epic,eepic}
\usepackage{bold-extra}
\usepackage{amsmath,amssymb}
\usepackage{pifont,wasysym}
\usepackage{tikz}

\newcommand{\abbrev}[1]{{\em #1}\xspace}
\newcommand{\ie}{\abbrev{i.e.}}
\newcommand{\eg}{\abbrev{e.g.}}
\newcommand{\vs}{\abbrev{vs.}}

\newcommand{\acronym}[1]{{\sc #1}\xspace}

\newcommand{\Cocoa}  {\acronym{Cocoa}}

\newcommand{\Cpu}    {\acronym{Cpu}}

\newcommand{\Ffi}    {\acronym{Ffi}}

\newcommand{\Gcc}    {\acronym{Gcc}}

\newcommand{\Hom}    {\acronym{Hom}}

\newcommand{\Iso}    {\acronym{Iso}}

\newcommand{\Posix}  {\acronym{Posix}}

\newcommand{\ProgLang}[1]{{\sc #1}\xspace}

\newcommand{\Awk}       {\ProgLang{Awk}}
\newcommand{\Cecil}     {\ProgLang{Cecil}}
\newcommand{\Cpp}       {\ProgLang{C{\small ++}}}
\newcommand{\Csharp}    {\ProgLang{C{\small\#}}}
\newcommand{\CL}        {\ProgLang{Common Lisp}}
\newcommand{\Clos}      {\ProgLang{Clos}}
\newcommand{\Cos}       {\ProgLang{Cos}}

\newcommand{\Dylan}     {\ProgLang{Dylan}}
\newcommand{\Eiffel}    {\ProgLang{Eiffel}}

\newcommand{\Fscript}   {\ProgLang{F-Script}}

\newcommand{\Java}      {\ProgLang{Java}}

\newcommand{\Objc}      {\ProgLang{Objective-C}}

\newcommand{\Perl}      {\ProgLang{Perl}}
\newcommand{\PerlSix}   {\ProgLang{Perl6}}

\newcommand{\Python}    {\ProgLang{Python}}

\newcommand{\Smalltalk} {\ProgLang{Smalltalk}}

\newcommand{\ttb}[1]{{\normalfont\ttfamily\bfseries #1}}
\newcommand{\opt}{\ensuremath{_{\text{opt}}}\xspace}
\newcommand{\lnk}{\ensuremath{\hookrightarrow}\xspace}

\newcommand{\hr}{\rule{\columnwidth}{0.33pt}}

\newcommand{\code}[1]{\lstinline[language=COS,style=samplecode]|#1|}
\newcommand{\objcode}[1]{\lstinline[language=OBJC,style=samplecode]|#1|}
\newcommand{\closcode}[1]{\lstinline[language=CLOS,style=samplecode]|#1|}

\newcommand{\eiffelcode}[1]{\lstinline[language=Eiffel,style=samplecode]|#1|}

\begin{document}

\lstdefinelanguage{CLOS}[]{Lisp}{}
\lstdefinelanguage{OBJC}[Objective]{C}{}
\lstdefinelanguage{COS}[ANSI]{C}{
  morekeywords={
    inline,
    classref,useclass,defclass,defproperty,endclass,makclass,
    genericref,usegeneric,defgeneric,makgeneric,defgenericv,makgenericv,
    defmethod,defalias,defnext,endmethod,retmethod,next_method,next_method_p,forward_message,
    self,self1,self2,self3,self4,self5,
    _1,_2,_3,_4,_5,_6,_7,_8,_9,
    _mth,_ret,_sel,_arg,
    test_assert,test_invariant,test_errno,
    RETVAL,
    PRE,POST,BODY,
    TRY,CATCH,CATCH_ANY,FINALLY,ENDTRY,THROW,RETHROW,
    PRT,OPRT,EPRT,UNPRT,
    NO,YES,
    OBJ,SEL,
    I8,I16,I32,I64,
    U8,U16,U32,U64,
    F64,C64,
    BOOL,STR,
    IMP1,IMP2,IMP3,IMP4,IMP5,
    Nil,True,False
  }
}

\lstdefinestyle{code}{
  basicstyle=\small\ttfamily,
  stringstyle=\normalfont,
  commentstyle=\normalfont\itshape,
  numberstyle=\tiny, 
  columns=fullflexible,
  xleftmargin=\parindent,
  escapeinside={//*}{*//}
}

\lstdefinestyle{samplecode}{
  style=code,
  emphstyle=\it,
  emph={obj,obj1,obj2,cls,cls1,cls2,spr,expr,str,func,file,line}
}

\lstdefinestyle{fltcode}{
  style=code,
  float=t,
  frame=t,
  captionpos=b,
  numbers=left
}


\lstnewenvironment{COS}[1][none]
  {\lstset{language=COS,style=code,numbers=#1}}
  {}
\lstnewenvironment{OBJC}[1][none]
  {\lstset{language=OBJC,style=code,numbers=#1}}
  {}
\lstnewenvironment{CLOS}[1][none]
  {\lstset{language=CLOS,style=code,numbers=#1}}
  {}
\lstnewenvironment{PERL}[1][none]
  {\lstset{language=Perl,style=code,numbers=#1}}
  {}
\lstnewenvironment{PYTHON}[1][none]
  {\lstset{language=Python,style=code,numbers=#1}}
  {}

\conferenceinfo{OOPSLA 2009}{October 25--29, 2009, Orlondo, Florida.}
\copyrightyear{2009}

\preprintfooter{Draft research paper for OOPSLA'09}

\title{The C Object System%
\titlenote{COS project:~{\tt http://sourceforge.net/projects/cos}}}
\subtitle{Using C as a High-Level Object-Oriented Language}
         
\authorinfo{Laurent Deniau}
{CERN -- European Organization for Nuclear Research}
{laurent.deniau@cern.ch}

\maketitle

\begin{abstract} 
The C Object System (\Cos) is a small C library which implements high-level concepts available in \Clos, \Objc and other object-oriented programming languages: {\em uniform object model} (class, meta\-class and property-metaclass), {\em generic functions, multi-methods, delegation, properties, exceptions, contracts} and {\em closures}. \Cos relies on the program\-mable capabilities of the C programming language to extend its syntax and to implement the aforementioned concepts as {\em first-class objects}. \Cos aims at satisfying several general principles like {\em simplicity, extensibility, reusability, efficiency} and {\em portability} which are rarely met in a single programming language. Its design is tuned to provide efficient and portable implementation of {\em message multi-dispatch} and {\em message multi-forwarding} which are the heart of code extensibility and reusability. With COS features in hand, software should become as flexible and extensible as with scripting languages and as efficient and portable as expected with C programming. Likewise, \Cos concepts should significantly simplify {\em adaptive} and {\em aspect-oriented} programming as well as {\em distributed} and {\em service-oriented} computing.
\end{abstract}
            
\category{D.3.3}
{C Programming Language}
{Language Constructs and Features}
\category{D.1.5}
{Programming Techniques}
{Object-oriented Programming}.

\terms
Object-oriented programming.
            
\keywords
Adaptive object model,
Aspects,
Class cluster,
Closure,
Contract,
Delegation,
Design pattern,
Exception,
Generic function,
Introspection,
High-order message,
Message forwarding,
Meta class,
Meta-object protocol,
Multimethod,
Open class model,
Predicate dispatch,
Programming language design,
Properties,
Uniform object model.
\section{Motivation\label{sec:why}}


The C Object System (\Cos) is a small framework which adds an {\em object-oriented layer} to the C programming language \cite{c99, harb02, hans97} using its {\em programmable capabilities}\footnote{In the sense of ``{\em Lisp is a programmable programming language}'', \cite{fod91}.}  while following the simplicity of \Objc \cite{objc91, objc08} and the extensibility of \Clos \cite{clos89,mop91,clos91}. \Cos aims to fulfill several general principles rarely met in a single programming language: {\em simplicity, extensibility, reusability, efficiency} and {\em portability}.


\subsection{Context}

\Cos has been developed in the hope to solve fundamental programming problems encountered in scientific computing and more specifically in {\em applied metrology} \cite{bos99, bos00}. Although this domain looks simple at first glance, it involves nonetheless numerous fields of computer science; from {\em low-level} tasks like the development of drivers, protocols or state machines, the control of hardware, the acquisition of data, the synchronization of concurrent processes, or the numerical analysis and modeling of huge data sets; to {\em high-level} tasks like the interaction with databases or web servers, the management of remote or distributed resources, the visualization of complex data sets or the interpretation of scripts to make the system configurable and controllable by non-programmers \cite{ems01, ems06, gio07}. Not to mention that scientific projects commonly have to rely on sparse human resources to develop and maintain for the long term such {\em continually-evolving-systems} (\ie R\&D). Therefore the challenge is ambitious but I firmly believe that \Cos provides the required features to simplify the development and the support of such systems as well as a wide variety of software projects.

\subsection{Principles}

Given the context, it is essential to reduce the multiplicity of the technologies involved, to simplify the development process, to enhance the productivity, to guarantee the extensibility and the portability of the code and to adapt the required skills to the available resources. Hence, the {\em qualities} of the programming language are essential for the success of such projects and should focus on the following principles:

\paragraph{Simplicity}

The language should be easy to learn and use. The training curve for an {\em average} programmer should be as short as possible what implies in particular a clear and concise syntax. Simplicity should become an asset which guarantees the quality of the code and allows writing complex constructions without being penalized by a complex formalism or by the multiplicity of the paradigms. \Cos can be learned within a few days by C programmers with some knowledge of object-oriented concepts, although exploiting the full power of \Cos requires some experience.

\paragraph{Extensibility}

The language should support the addition of new features or the improvement of existing features without changing significantly the code or the software architecture. Concepts like {\em polymorphism, message dispatch} and {\em open class model} help to achieve good {\em flexibility} and {\em extensibility} by reducing coupling. But they usually have a strong impact on the {\em efficiency}. \Cos dispatches messages with an efficiency in the range of the \Cpp virtual member functions. 

\paragraph{Reusability}

The language should support code reusability, namely the ability to reuse or quickly adapt existing components to unforeseen tasks. It is easier to achieve this goal if the language allows writing {\em generic code}, either by parameterization, either by abstraction, to ease the componentization of design patterns \cite{rege99,meyer06-1,meyer06-2}. To support the development of {\em generic components}, \Cos provides multi-methods to handle dynamic and polymorphic {\em collaboration} and delegation to handle dynamic and polymorphic {\em composition}.

\paragraph{Efficiency}

A general purpose programming language must be efficient, that is it must be able to translate all kinds of algorithms into programs running with {\em predictable} resource usage (mainly \Cpu and memory) consistent with the processes carried out. In this respect, programming languages with an abstract machine close to the physical machine ---~a {\em low-level} language~--- offer generally better results. C is admittedly known to achieve good efficiency.

\paragraph{Portability}

A general purpose programming language must be portable, that is it must be widely available on many architectures and it must be accessible from almost any other languages (\Ffi). This point often neglected brings many advantages: it improves the software reliability, it reduces the deployment cost, it enlarges the field of potential users and it helps to find trained programmers. Regarding this point, {\em normalized} programming languages (\Iso) get the advantage. ISO C89 is normalized and well known for its availability and portability.

\subsection{Proposition}

\Cos extends the C programming language with concepts \cite{mitch01} mostly borrowed from \Objc and \Clos. The choice of designing the language as a C library instead of a compiler allowed to quickly explore various object models, but R.E.~Johnson's paper on the dynamic object model \cite{john98} definitely focused my research towards the final design:

{\em ``If a system is continually changing, or if you want users to be able to extend it, then the Dynamic Object Model architecture is often useful. [...] Systems based on Dynamic Object Models can be much smaller than alternatives. [...] I am working on replacing a system with several millions lines of code with a system based on a dynamic object model that I predict will require about 20,000 lines of code. [...] This makes these systems easier to change by experts, and (in theory) should make them easier to understand and maintain. But a Dynamic Object Model is hard to build. [...] A system based on a Dynamic Object Model is an interpreter, and can be slow.''}.

This adaptive object model \cite{john00, john02} is actually what \Cos provides, but at the level of the C programming languages without significant efficiency loss. In particular, \Cos has been designed to support efficiently two key concepts ---~{\em multi-methods} and {\em fast generic delegation}~--- and provides a {\em uniform object model} where classes, generics and methods are {\em first-class objects}. Incidentally, \Cos strengthens inherently all the guidelines stated in \cite{bos01} to build ``{\em flexible, usable and reusable object-oriented frameworks}'' as well as architectural pattern proposed in \cite{tel05} to design {\em flexible component-based frameworks}.


\section{Overview\label{sec:cos}}

\Cos is a small framework entirely written in portable\footnote{Namely C89 and C99 variadic macros.} C99 which provides programming paradigms like {\em objects, classes, metaclasses, generic functions, multi-methods, delegation, properties, exceptions, contracts} and {\em closures}. \Cos syntax and features are directly available at the C source code level through the use of the language keywords defined in the header file \code{<cos/Object.h>}.


\subsection{Concepts}

\paragraph{Polymorphism}

This concept available in {\em object-oriented} programming languages is the heart of software extensibility because it postpones to runtime the resolution of methods invocation and reduces coupling between callers and callees. Besides, if the polymorphic types are dynamic, the coupling becomes almost inexistent and code size and complexity are significantly reduced. On one hand, these simplifications usually {\em improve the programmer understanding} who makes less conceptual errors, draws simpler designs and increases its productivity. On the other hand, dynamic typing postpones the detection of unknown messages at runtime, with the risk to see programs ending prematurely. But well tested software reduce this risk to exceptional situations. 





\paragraph{Collaboration}

Software development is mainly about building collaborations between entities, namely objects. As soon as polymorphic objects are involved everywhere to ensure good software extensibility and reusability, one needs {\em polymorphic collaboration} implemented by {\em multi-methods}. They reduce strong coupling that exist in the Visitor pattern (or equivalent) as well as the amount of code needed to achieve the task. \Cos provides message multi-dispatch with an efficiency in the range of the \Cpp virtual member function.

\paragraph{Composition}

The composition of objects and behaviors is a well known key-concept in software design. It enhances software flexibility by introducing levels of indirection in objects and behaviors. Most structural and behavioral design patterns described in \cite{gof95} introduce such indirections, but at the price of an increased code complexity and coupling and hence a decreased reusability of the components built. The {\em delegation} is an effective mechanism which allows managing the composition of both, objects and behaviors, without introducing coupling. \Cos provides delegation with the efficiency of message dispatch, {\em seemingly a unique feature}.

\paragraph{Reflection}

Reflection is a powerful aspect of adaptive object models which, amongst others, allows to mimic the behavior of interpreters. \Cos provides full introspection and limited intercession on polymorphic types and behaviors, that is classes, generics and methods, as well as object attributes through the definition of properties. Since all \Cos components are {\em first-class objects}, it is trivial to replace creational patterns \cite{gof95} by generic functions (section \ref{ssec:spat}).

\paragraph{Encapsulation}

Encapsulation is a major concern when developing libraries and large-scale software. \Cos enforces encapsulation of class implementation because encapsulation is not only a matter of managing coupling but also a design issue. Besides, the object behaviors are represented by generics which favors the {\em separation of concerns} of interfaces and reduces cross-interfaces dependencies \cite{bos01}. Moreover, the open class model of \Cos allows extending classes {\em on need} without breaking the encapsulation (\ie without ``{\em reopening the box}'') and reduces the risk  of premature design.



\paragraph{Ownership}

The management of object life cycles requires a clear policy of ownership and scope rules. In languages like C and \Cpp where semantic {\em by value} prevails, the burden is put on the programmer's shoulders. In languages like \Java, \Csharp and D where semantic {\em by reference} prevails, the burden is put on the garbage collector. In this domain, \Cos lets the developer choose between garbage collection (\eg Boehm GC \cite{boehm02}) and {\em manual reference counting with rich semantic} (section~\ref{ssec:obj}).

\paragraph{Concurrency}

\Cos has been designed from the beginning with concurrency in mind and shares only its dictionary of {\em static components}. Per thread resources like {\em message caches} and {\em autorelease pools} rely on either thread-local-storage or thread-specific-key according to the availability.



\subsection{Components}

\begin{figure}\hr
\begin{center}
\vspace{-3mm}\hspace{-3mm}
\includegraphics[width=0.85\columnwidth,keepaspectratio=true]{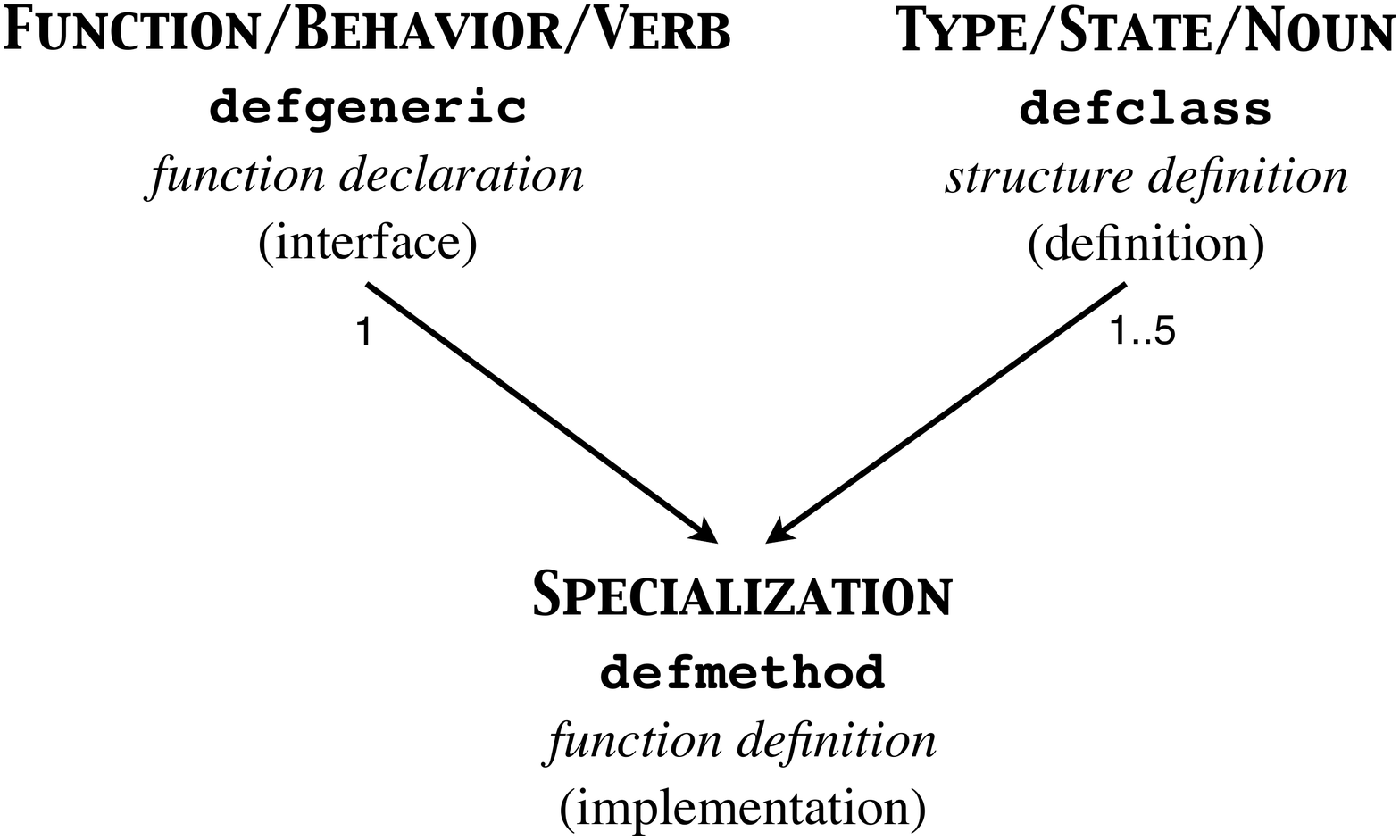}
\vspace{-4mm}
\end{center}
\caption{Roles of \Cos components and their equivalent C-forms. {\em Multi-methods are classes specialization of generics}.\label{fig:cgm}}
\end{figure}

The object-oriented layer of \Cos is based on three components (figure~\ref{fig:cgm}) borrowed from \Clos which characterize the {\em open object model} described in depth in \cite{clos89} and \cite{mop91}.

\paragraph{Classes}

Classes play the same role as {\em structures} in C and define object {\em attributes}. They are bound to their {\em superclass} and {\em metaclasses} and define supertypes-subtypes hierarchies.

\paragraph{Generics}

Generics play the same role as {\em function declarations} in C and define {\em messages}. They are essential actors of code extensibility and ensure correctness of formal parameters of messages between callers and callees.

\paragraph{Methods}

Methods play the same role as {\em function definitions} in C and define {\em specializations} of generics. A method is invoked if the message belongs to its generic and the receivers match its classes (multi-methods).

\mbox{}\newline
The similarities between \Cos components and their equivalent C-forms let C programmers with some notions of object-oriented design be productive rapidly. The open object model allows defining components in different places and therefore requires an extra linking iteration to collect their {\em external symbols}: link~$\rightarrow$~collect\footnote{\Cos mangled symbols are collected with the {\tt nm} command or equivalent.}~$\rightarrow$~re-link. This fast iteration is automatically performed by the makefiles coming with \Cos before the final compilation stage that builds the executable or the dynamic library.

\subsection{Syntax}

\Cos introduces new keywords to extend the C language with a user-friendly syntax half-way between \Objc and \Clos. {\em\Cos parses its syntax and generates code with the help of its functional C preprocessing library}\footnote{The description of this \code{cpp} library is beyond the scope of this paper.}; a module of a few hundred C macros which was developed for this purpose. It offers limited parsing capabilities, token recognition, token manipulation and algorithms like {\em eval, map, filter, fold, scan, split} borrowed from functional languages and working on tuples of tokens.
As a rule of thumb, all \Cos symbols and macros are mangled to avoid unexpected collisions with other libraries, including keywords which can be disabled. 

Despite of its dynamic nature, \Cos tries hard to detect all syntax errors, type errors and other mistakes at compile-time by using {\em static asserts} or similar tricks and to emit meaningful diagnostics. The only point that \Cos cannot check at compile time is the understanding of a message by the receivers; an important ``feature'' to reduce coupling.

The syntax and grammar of \Cos are summarized in the figures~\ref{fig:clsgram}, \ref{fig:gengram}, \ref{fig:mthgram}, \ref{fig:ctrgram}, \ref{fig:prpgram} and \ref{fig:exgram}, following the notation of the C99 standard \cite{c99}.


\section{Classes {\em (nouns)}\label{sec:cls}}

\Cos allows defining and using classes as easily as in other object-oriented programming languages.

\subsection{Using classes}

The \code{useclass()} {\em declaration} allows accessing to classes as {\em first-class objects}. The following simple program highlights the similarities between \Cos and \Objc:
\begin{COS}[left]
#include <cos/Object.h>
#include <cos/generics.h>//*\label{lst:loadgen}*//

useclass(Counter, (Stdout)out);//*\label{lst:usecls}*//

int main(void) {
  OBJ cnt = gnew(Counter);//*\label{lst:obj}*//
  gput(out,cnt);//*\label{lst:put}*//
  gdelete(cnt);//*\label{lst:end}*//
}
\end{COS}
which can be translated line-by-line into \Objc by:
\begin{OBJC}[left]
#include <objc/Object.h>
// Counter interface isn't exposed intentionally

@class Counter, Stdout;

int main(void) {
  id cnt = [Counter new];
  [Stdout put: cnt];
  [cnt release];
}
\end{OBJC}
Line~\ref{lst:loadgen} makes the standard generics like \code{gnew}, \code{gput} and \code{gdelete}\footnote{By convention, the name of generics always starts by a '\code{g}'.} visible in the current translation unit. \Objc doesn't need this information since methods are bound to their class, but if the user wants to be warned for incorrect use of messages, the class definition must be visible. This example shows that \Cos requires less information than \Objc to handle compile-time checks what leads to better code insulation and reduces useless recompilations. Moreover, it offers fine tuning of exposure of interfaces since only the used generic functions have to be visible.

Line~\ref{lst:usecls} declares the class \code{Counter}\footnote{By convention, the name of classes always starts by an uppercase letter.}  and the alias \code{out} for local replacement of the class \code{Stdout}, both classes being supposedly defined elsewhere.
In line~\ref{lst:obj}, the generic type \code{OBJ} is equivalent to \objcode{id} in \Objc.

\noindent
Lines~\ref{lst:obj}~--~\ref{lst:end} show the life cycle of objects, starting with \code{gnew} (resp. \code{new}) and ending with \code{gdelete} (resp. \code{release}). They also show that generics {\em are} functions (\eg one can take their address). Finally, the line~\ref{lst:put} shows an example of multi-method where the message \code{gput(_,_)} will look for the specialization \code{gput(mStdout,Counter)} whose meaning is discussed in section~\ref{sec:mth}. In order to achieve the same task, \Objc code has to rely on the Visitor pattern, a burden that requires more coding, creates static dependencies (strong coupling) and is difficult to extend.

\begin{figure}\hr
\begin{center}
{\em
\begin{tabbing}
xxx\= xxx\= \hspace*{17em}\= \hspace*{2em} \= \kill
class-declaration: \\
\> \ttb{useclass(} class-decl-list \ttb{);} \\
\\
class-decl-list: \\
\> class-decl \\
\> class-decl-list \ttb{,} class-decl \\
\\
class-decl: \\
\> class-name \\
\> \ttb{(} class-name \ttb{)} local-name \\
\\
class-definition: \\
\> \ttb{defclass(} class-specifier \ttb{)} \\ \lnk
\> \> struct-declaration-list		\>\> (c99) \\ \lnk
\> \ttb{endclass} \\
\\
class-instantiation: \\
\> \ttb{makclass(} class-specifier \ttb{);} \\
\\
class-specifier: \\
\> class-name \\
\> class-name \ttb{,} \rule{1.5ex}{1pt} \>\> (root class) \\
\> class-name \ttb{,} superclass-name \\
\\
\{class, superclass, local\}-name: \\
\> identifier					\>\>\> (c99)
\end{tabbing}
}
\end{center}
\caption{Syntax summary of classes.\label{fig:clsgram}}
\end{figure}

\subsection{Defining classes}

The definition of a class is very similar to a C structure:
\begin{COS}
defclass(Counter)
  int cnt;
endclass
\end{COS}
which is translated in \Objc as:
\begin{OBJC}
@interface Counter : Object {
  int cnt;
}
// declaration of Counter methods not shown
@end
\end{OBJC}
or equivalently in \Clos as:
\begin{CLOS}
(defclass Counter (Object) ((cnt)) )
\end{CLOS}
The \code{Counter} class derives from the root class \code{Object} --- the default behavior when the superclass isn't specified --- and defines the attribute \code{cnt}.

\paragraph{Class visibility\label{par:new}}

{\em What must be visible and when?} In order to manage coupling, \Cos provides three levels of visibility: none, declaration and definition. If you only use the generic type \code{OBJ}, nothing is required {\em (no coupling)}:
\begin{COS}
OBJ gnew(OBJ cls) {
  return ginit(galloc(cls));
}
\end{COS}
If you want to create instances of a class, only the declaration is required {\em (weak coupling)}:
\begin{COS}
OBJ gnewBook(void) {
  useclass(Book); // local declaration
  return gnew(Book);
}
\end{COS}
If you want to define subclasses, methods or instances with automatic storage duration, the class definition must be visible {\em (strong coupling)}.

\subsection{Class inheritance}

\begin{figure}\hr
\begin{center}
\includegraphics[width=\columnwidth,keepaspectratio=true]{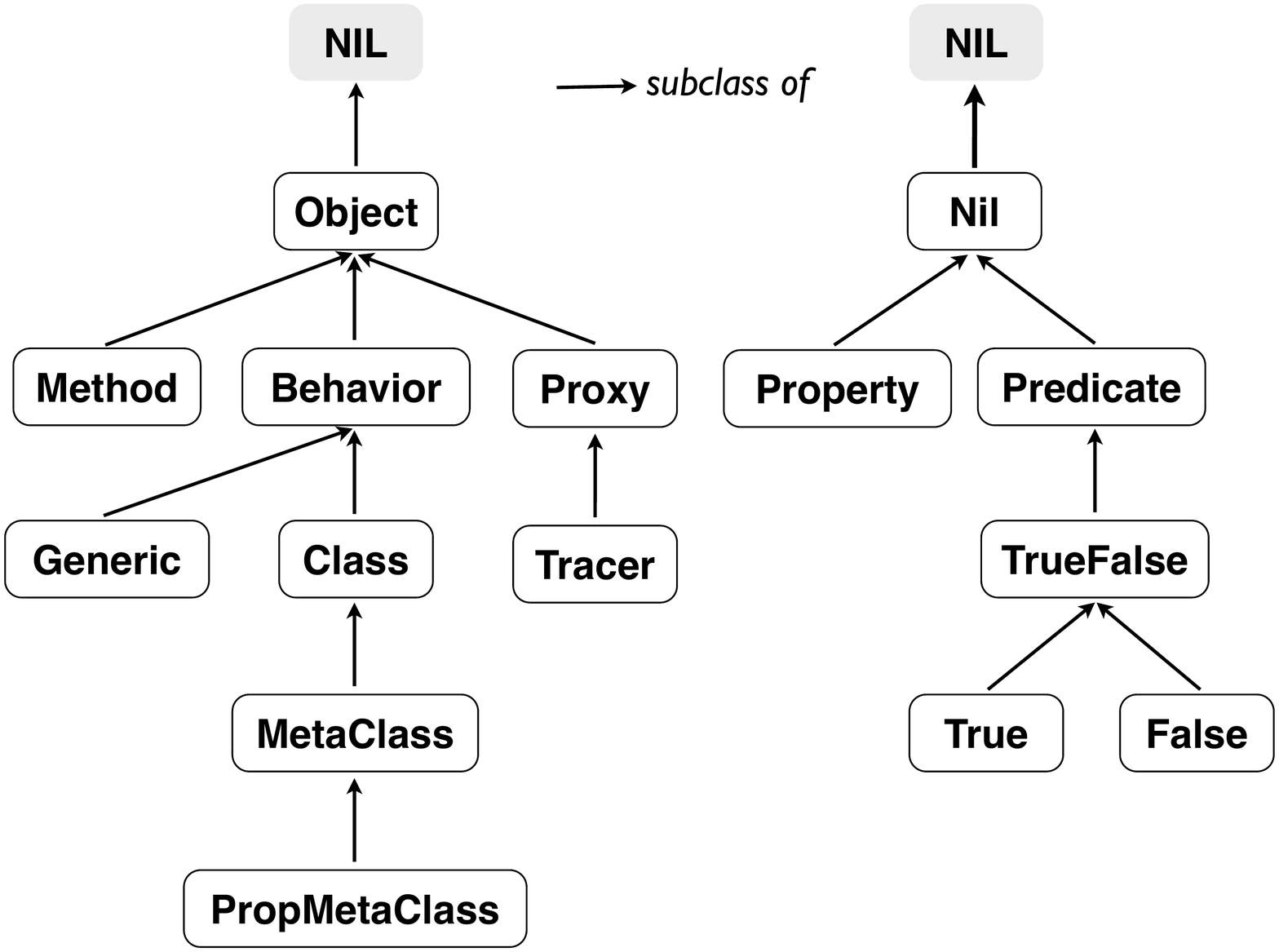}
\end{center}
\vspace{-1mm}
\caption{Subset of \Cos core classes hierarchy.\label{fig:clstree}}
\end{figure}

Class inheritance is as easy in \Cos as in other object-oriented programming languages. Figure~\ref{fig:clstree} shows the hierarchies of the core classes of \Cos deriving from the root classes \code{Object} and \code{Nil}. As an example, the \code{MilliCounter} class defined hereafter derives from the class \code{Counter} to extend its resolution to thousandths of count:
\begin{COS}
defclass(MilliCounter, Counter)
  int mcnt;
endclass
\end{COS}
which gives in \Objc:
\begin{OBJC}
@interface MilliCounter : Counter {
  int mcnt;
}
// declaration of MilliCounter methods not shown
@end
\end{OBJC}
and in \Clos:
\begin{CLOS}
(defclass MilliCounter (Counter) ((mcnt)) )
\end{CLOS}
In the three cases, the derived class inherits the attributes and the methods of its superclass.
Since \Cos aims at insulating classes as much as possible, it discourages direct access to superclass attributes by introducing a syntactic indirection which forces the user to write \code{obj->}{\tt\em Super.attribute} instead of \code{obj->}{\tt\em attribute}. The inheritance of {\em multi-methods} has a different meaning and will be discussed in section~\ref{sec:mth}.

\paragraph{Root class}

Defining a root class is an exceptional task but it may be a necessity in some rare cases. \Cos uses the terminal symbol $\perp$\footnote{$\perp$ means ``{\em end of hierarchy}'' or \code{NIL}, but not the class \code{Nil}.} (represented by '\code{_}') to declare a class as a root class. For example, \code{Object} is an important root class with the following simple definition:
\begin{COS}
defclass(Object,_)
  U32 id;		// object's class identity
  U32 rc;		// reference counting
endclass
\end{COS}
But its methods must be defined with care since they provide all the essential functionalities inherited by other classes.

\paragraph{Class rank}

\Cos computes at compile-time the inheritance depth of each class. The rank of a root class is zero (by definition) and each successive subclass increases the rank.

\paragraph{Dynamic inheritance}

\Cos provides the message \code{gchange}\-\code{Class}\code{(obj,cls)} to change the class of \code{obj} to \code{cls} iff it is a superclass of \code{obj}'s class; and the message \code{gunsafe}\-\code{Change}\-\code{Class(obj,cls,spr)} to change the class of \code{obj} to \code{cls} iff both classes share a common superclass \code{spr} and the instance size of \code{cls} is lesser or equal to the size of \code{obj}. These messages are useful for implementing {\em class clusters}, {\em state machines} and {\em adaptive behaviors}.

\subsection{Meta classes\label{ssec:meta}}

\begin{figure}\hr
\begin{center}
\vspace*{-3mm}\hspace*{-7mm}
\includegraphics[width=1.15\columnwidth,keepaspectratio=true]{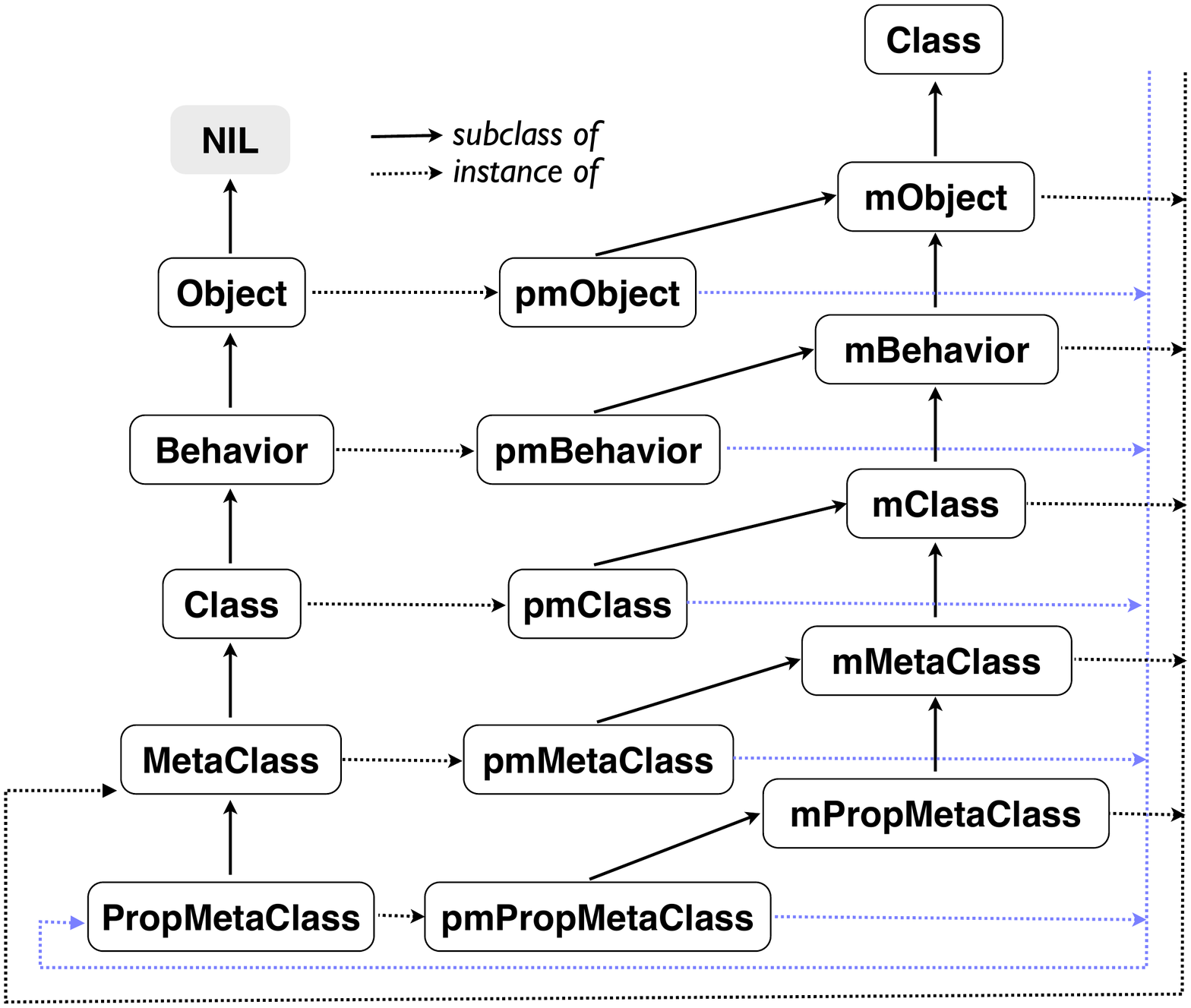}
\end{center}
\caption{\Cos core classes hierarchy with metaclasses.\label{fig:metatree}}
\end{figure}

Like in \Objc, a \Cos class definition creates a parallel hierarchy of metaclass which facilitates the use of {\em classes as first-class objects}. Figure~\ref{fig:metatree} shows the complete hierarchy of the \code{PropMetaClass} class, including its metaclasses.

\paragraph{Class metaclass}

The metaclasses are {\em classes of classes} implicitly defined in \Cos to ensure the coherency of the type system: to each class must correspond a metaclass \cite{raza04}. Both inheritance trees are built in parallel: if a class \code{A} derives from a class \code{B}, then its metaclass \code{mA}\footnote{The metaclass name is always the class name prefixed by a '\code{m}'.} derives from the metaclass \code{mB} --- except the root classes which derive from \code{NIL} and have their metaclasses deriving from \code{Class} to {\em close the inheritance path}. Metaclasses are instances of the class \code{MetaClass}.

\paragraph{Property metaclass}

In some design patterns like Singleton or Class Cluster, or during class initialization (section \ref{par:clsini}), the automatic derivation of the class metaclass from its superclass metaclass can be problematic as detailed in \cite{bour98}. To solve the problem \Cos associates to each class a {\em property metaclass which cannot be derived}; that is all methods specialized on the property metaclass can only be reached by the class itself. In order to preserve the consistency of the hierarchy, a property metaclass must always derive from its class metaclass, namely \code{pmA}\footnote{The property metaclass name is always the class name prefixed by a '\code{pm}'.}  (resp. \code{pmB}) derives from \code{mA} (resp. \code{mB}) as shown in the figure~\ref{fig:metatree}. Property metaclasses are instances of the class \code{PropMetaClass}.

\paragraph{Class objects}

\begin{figure}\hr
\begin{center}
\vspace*{-2mm}
\includegraphics[width=0.9\columnwidth,keepaspectratio=true]{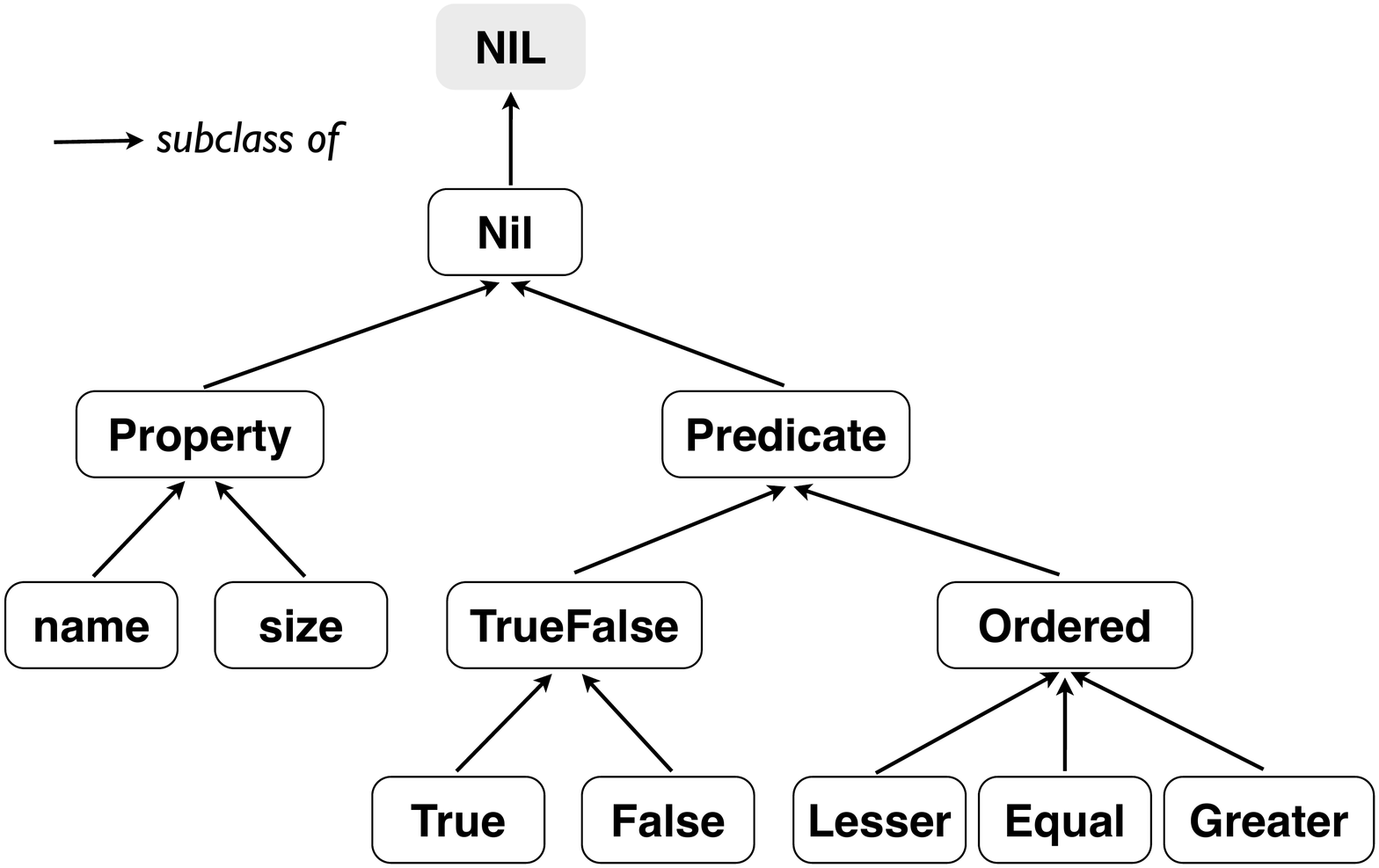}
\vspace*{-2mm}
\end{center}
\caption{Subset of \Cos core class-predicates hierarchy.\label{fig:predtree}}
\end{figure}

With multi-methods and metaclasses in hands, it is possible to use classes as common objects. Figure~\ref{fig:predtree} shows the hierarchy of the core class-objects used in \Cos to specialized multi-methods with specific {\em states}. For instance messages like \code{gand}, \code{gor} and \code{gnot} are able to respond to messages containing the class-predicates \code{True}, \code{False} and \code{TrueFalse}. The root class \code{Nil} is a special class-object which means {\em no-object} but still safe for message dispatch: sending a message to \code{Nil} is safe, but not to \code{NIL}.

\paragraph{Type system}

The \Cos type system follows the rules of \Objc, that is polymorphic objects have opaque types (ADT) outside their methods and are {\em statically and strongly typed inside}; not to mention that multi-methods reduce significantly the need for runtime identification of polymorphic parameters. Furthermore, the set of {\em class~--~metaclass~--~property-metaclass} forms a coherent hierarchy of classes and types which offers better consistency and more flexibility than in \Objc and \Smalltalk where metaclasses are not explicit and derive directly from \code{Object}.

\subsection{Class instances\label{ssec:obj}}

\paragraph{Object life cycle}

The life cycle of objects in \Cos is very similar to other object-oriented programming languages, namely it {\em starts} by creation (\code{galloc}) followed by initialization (\code{ginit} and variants) and {\em ends} with deinitialization (\code{gdeinit}) followed by destruction (\code{gdealloc}). In between, the user manages the ownership of objects (\ie dynamic scope) with \code{gretain}, \code{grelease} and \code{gautoRelease} like in \Objc. The {\em copy initializer} is the specialization of the generic \code{ginitWith(_,_)} for the same class twice. The {\em designated initializer} is the initializer with the most coverage which invokes the designated initializer of the superclass using \code{next_method}. Other initializers are {\em secondary initializers} which must invoke the designated initializer \cite{cocoa07}.

\paragraph{Object type}

In \Cos (resp. \Objc), objects are always of dynamic type because the type of \code{galloc} (resp. \objcode{alloc}) is \code{OBJ} (resp. \objcode{id}). Since it is the first step of the life cycle of objects in both languages, the type of objects can never be known statically, except inside their own multi-methods. That is why \Cos (resp. \Objc) provides the message \code{gisKindOf(obj,cls)} (resp. \objcode{[obj isKindOf: cls]}) to inspect the type of objects. But even so, it would be dangerous to use a static cast to convert an object into its expected type because dynamic design patterns like Class Cluster and Proxy might override \code{gisKindOf} for their use. \Cos also provides the message \code{gclass(obj)} which returns \code{obj}'s class.

\paragraph{Object identity}

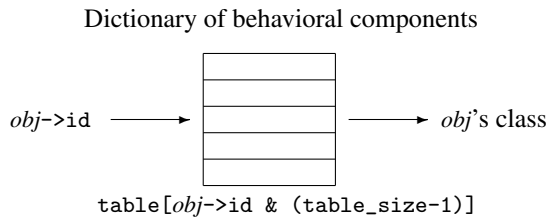
\begin{figure}\hr
\begin{center}
\begin{picture}(200,80)(5,0)
\put(2,32){\code{obj->id}}
\put(40,35){\vector(1,0){30}}

\put(30,70){Dictionary of behavioral components}
\put(75,10){\grid(50,50)(50,10)}
\put(35,0){\code{table[obj->id & (table_size-1)]}}

\put(130,35){\vector(1,0){30}}
\put(165,32){\code{obj}'s class}
\end{picture}
\end{center}
\caption{Lookup to retrieve object's class from object's id.\label{fig:clslut}}
\end{figure}

In \Cos, an object is bounded to its class through a unique $32$-bit identifier produced by a linear congruential generator which is also a generator of the cyclic groups $\mathbb{N}/2^k\mathbb{N}$ for $k=2..32$. This powerful algebraic property allows retrieving efficiently the class of an object from the components table using its identifier as an index (Figure~\ref{fig:clslut}). Comparing to pointer-based implementations, the unique identifier has four advantages:

\vspace{1ex}\noindent
\fbox{\parbox{0.97\columnwidth}{{\em It ensures better behavior of cache lookups under heavy load (uniform hash), it makes the hash functions very fast (sum of shifted {\em\code{id}}s), it is smaller than pointers on 64-bit}}}

\vspace{1ex}\noindent
\fbox{\parbox{0.97\columnwidth}{{\em
machines and it can store extra information (high bits) like class ranks to speedup linear lookup in class hierarchies}.}}
\vspace{-0.15ex}
 

\paragraph{Automatic objects}

Since \Cos adds an object-oriented layer on top of the C programming language, it is possible to create objects with automatic storage duration (\eg on the stack) using compound literals (C99). In order to achieve this, the class definition must be visible and the developer of the class must provide a special constructor. For example the constructor \code{aStr(''a string'')}\footnote{By convention, {\em automatic} constructors always starts by an \code{'a'}.} is equivalent to the \Objc directive \objcode{@''a string''}. \Cos already provides automatic constructors for many common objects like \code{Char}, \code{Short}, \code{Int}, \code{Long}, \code{Float}, \code{Complex}, \code{Range}, \code{Functor} and \code{Array}. {\em Automatic constructors allow creating efficiently temporary objects with local scope and enhance the flexibility of multi-methods}. For example, the initializer \code{ginitWith(_,_)} and its variants can be used in conjunction with almost all the automatic constructors aforementioned. Thanks to the rich semantic of \Cos reference counting, if an automatic object receives the message \code{gretain} or \code{gautoDelete}, it is automatically cloned using the message \code{gclone} and the new copy with dynamic scope is returned.

\paragraph{Static objects}

Static objects can be built in the same way as automatic objects except that they require some care in multi-threaded environments. It is worth to note that all \Cos{} {\em components} have static storage duration and consequently are {\em insensitive to ownership} and cannot be destroyed.

\subsection{Implementing classes}

Class instantiations create the {\em class} objects using the keyword \code{makclass} and the same {\em class-specifier} as the corresponding \code{defclass}. \Cos checks at compile-time if both definitions match. The counters implementation follows:
\begin{COS}
makclass(Counter);
makclass(MilliCounter,Counter);
\end{COS}
which is equivalent in \Objc to:
\begin{OBJC}
@implementation Counter
// definition of Counter methods not shown
@end
@implementation MilliCounter : Counter
// definition of MilliCounter methods not shown
@end
\end{OBJC}






\paragraph{Class initialization\label{par:clsini}}

For the purpose of pre-initialization, \Cos ensures to invoke {\em once by ascending class rank} (superclass first) all specializations of the message \code{ginitialize} on {\em property} metaclass before the first message is sent. Likewise, \Cos ensures to invoke {\em once by descending class rank} (subclasses first) all specializations of the message \code{gdeinitialize} on {\em property} metaclass after exiting \code{main}.

\section{Generics {\em (verbs)}\label{sec:gen}}

We have already seen in previous code samples that generics can be used as functions. But generics take in fact multiple forms and define each:

\begin{itemize}
\item a {\em function declaration} (\code{defgeneric}) which ensures the correctness of the signature of its methods (\code{defmethod}), aliases (\code{defalias}) and next-methods (\code{defnext}).

\item a {\em function definition} used to dispatch the message and to find the most specialized method belonging to the generic and matching the classes of the {\em receivers}.

\item an {\em object} holding the generic's metadata: {\em the selector}.
\end{itemize}
A generic function has one definition of its semantics and is, in effect, a verb raised at the same level of abstraction as a noun \cite{fod91}. Figure~\ref{fig:gengram} summarizes the syntax of generics, half way between the syntax of generic's definition in \Clos and the syntax of method's declaration in \Objc.

\begin{figure}\hr
\begin{center}
{\em\vspace{-1mm}
\begin{tabbing}
xxx\= xxx\= \hspace*{17.5em}\= \hspace*{1.5em}\= \kill
generic-declaration: \\
\> \ttb{usegeneric(} generic-decl-list \ttb{);} \\
\\
generic-decl-list: \\
\> generic-decl \\
\> generic-decl-list \ttb{,} generic-decl \\
\\
generic-decl: \\
\> generic-name \\
\> \ttb{(} generic-name \ttb{)} local-name \\
\\
generic-definition: \\
\> \ttb{defgeneric(} generic-specifier \ttb{);} \\
\\
generic-variadic-definition: \\
\> \ttb{defgenericv(} generic-specifier \ttb{, ...);} \\
\\
generic-specifier: \\
\> return-type \ttb{,} generic-def \ttb{,} param-list \\
\\
generic-def: \\
\> generic-name \\
\> \ttb{(} class-name \ttb{)} generic-name \\
\\
param-list: \\
\> param-decl \\
\> param-list \ttb{,} param-decl \\
\\
param-decl: \\
\> param-name\opt 					\>\> (selector) \\
\> \ttb{(} param-type \ttb{)} param-name \\
\\
\{return, param\}-type: \\
\> type-name							\>\>\> (c99) \\
\\
\{generic, param\}-name: \\
\> identifier						\>\>\> (c99)
\end{tabbing}
\vspace{-1mm}}
\end{center}
\caption{Syntax summary of generics.\label{fig:gengram}}
\end{figure}

\paragraph{Generic rank}

The rank of a generic is the number of receivers in its {\em param-list}. \Cos supports generics from rank $1$ to $5$ what should be enough in practice since rank 1 to 4 already cover all the multi-methods defined in the libraries of \Cecil and \Dylan \cite{duj98,zib02,cecil04}.

\subsection{Message dispatch}

\Cos dispatch uses global caches (one per generics rank) implemented with hash tables to speedup method lookups. The caches solve slot collisions by growing until they reach a configurable upper bound of slots. After that, they use packed linked list incrementally built to hold a maximum of $3$ cells. Above this length, the caches start to forget cached methods --- a required behavior when dynamic class creation is supported. The lookup uses {\em fast asymmetric hash functions} (sum of shifted \code{id}s) to compute the cache slots and ensures uniform distribution even when all selectors have the same type or specializations on permutations exist.




\paragraph{Fast messages}

\Cos lookup is simple enough to allow some code inlining on the caller side to speedup message dispatch. Fast lookup is enabled {\em up to the generic rank} specified by \code{COS_FAST_MESSAGE} --- from disabled (0) to all (5, default) --- before the generic definitions (\code{defgeneric}).

\subsection{Declaring generics}

Generic declarations are less common than class declarations but they can be useful when one wants to use generics as first-class objects. Since generic definitions are more often visible than class definitions, it is common to rename them locally as in the following short example:
\begin{COS}
void safe_print(OBJ obj) {
  usegeneric( (gprint) prn );
  if ( gunderstandMessage1(obj, prn) == True )
    gprint(obj);
}
\end{COS}
which gives in \Objc:
\begin{OBJC}
void safe_print(id obj) {
  SEL prn = @selector(print);
  if ( [obj respondsToSelector: prn] == YES )
    [obj print];
}
\end{OBJC}

\subsection{Defining generics\label{ssec:gen}}

Definitions of generics correspond to function declarations in C and differ from \Objc method declarations by the fact that they are neither bound to classes (prefix '\objcode{-}') nor to metaclasses (prefix '\objcode{+}'). The following definitions:
\begin{COS}
defgeneric(void, gincr, _1);            //*\hfill*// // rank 1
defgeneric(void, gincrBy, _1, (int)by); //*\hfill*// // rank 1
defgeneric(OBJ , ginitWith, _1, _2);    //*\hfill*// // rank 2
defgeneric(OBJ , ggetAt, _1, at);       //*\hfill*// // rank 2
defgeneric(void, gputAt, _1, at, what); //*\hfill*// // rank 3
\end{COS}
can be translated into \Clos as:
\begin{CLOS}
(defgeneric incr (obj))
(defgeneric incr-by (obj by))
(defgeneric init-with (obj with))
(defgeneric get-at (obj at))
(defgeneric put-at (obj at what))
\end{CLOS}
Selector parameters like \code{at} are called {\em open types} (no parenthesis) since their type can vary for each specialization. Other parameters like \code{by} are called {\em closed types} (with parenthesis) and have fixed types and names: specializations must use the same types and names as defined by the generic. This enforces the semantic of {\em monomorphic} parameters which could be ambiguous otherwise: \code{int offset} \vs \code{int index}.





\section{Methods\label{sec:mth}}

Methods are defined using a similar syntax as generics as summarized in figure~\ref{fig:mthgram}. The following code defines a method specialization of the generic \code{gincr} for the class \code{Counter}:
\begin{COS}
defmethod(void, gincr, Counter)
  self->cnt++;
endmethod
\end{COS}
which in \Objc gives (within \objcode{@implementation}):
\begin{OBJC}
- (id) incr {
  self->cnt++;
}
\end{OBJC}




\paragraph{Methods specializers}

The {\em receivers} can be equivalently accessed through \code{self}{\em n}\footnote{\code{self} and \code{self1} are equivalent.} whose types correspond to their class specialization (\eg \code{struct} \code{Counter*}) and through unnamed parameters \code{_}{\em n} whose types are \code{OBJ} for $1\leq n\leq g$, where $g$ is the rank of the generic. It is important to understand that \code{self}{\em n} and \code{_}{\em n} are bound to the same object, but \code{self}{\em n} provides a statically typed access which allows {\em treating \Cos objects like normal C structures}.

\paragraph{Multi-methods}

Multi-methods are methods with more than one receiver and do not require special attention in \Cos. The following example defines the assign-sum operator (\ie~\code{+=}) specializations which adds 2 or 3 \code{Counter}s:
\begin{COS}
defmethod(OBJ, gaddTo, Counter, Counter)
  self->cnt += self2->cnt;
  retmethod(_1); // return self
endmethod

defmethod(OBJ, gaddTo2, Counter,Counter,Counter)
  self->cnt += self2->cnt + self3->cnt;
  retmethod(_1); // return self
endmethod
\end{COS}

\vspace{1ex}\noindent
\fbox{\parbox{\columnwidth}{{\em
About half of \Cos generics have a rank $> 1$ (multi-methods) and cover more than $80\%$ of all the methods specializations.}}}
\vspace{1ex}




\paragraph{Class methods}

Class methods are methods specialized for classes deriving from \code{Class} what includes all metaclasses:
\begin{COS}
defmethod(void, ginitialize, pmMyClass)
  // do some initialization specific to MyClass.
endmethod

defmethod(OBJ, gand, mTrue, mFalse)
  retmethod(False); // return the class-object False
endmethod
\end{COS}

\paragraph{Method aliases}

\Cos allows specializing compatible generics with the same implementation. The following aliases define specializations for \code{gpush}, \code{gtop} and \code{gpop} which share the specializations of \code{gput}, \code{gget} and \code{gdrop} respectively:
\begin{COS}
defalias(void, (gput )gpush, Stack, Object);
defalias(OBJ , (gget )gtop , Stack, Object);
defalias(void, (gdrop)gpop , Stack, Object);
\end{COS}

\begin{figure}\hr
\begin{center}
{\em\vspace{-1mm}
\begin{tabbing}
xxx\= xxx\= \hspace*{14.5em}\= \hspace*{2.6em}\= \hspace*{1.9em}\= \kill
method-definition: \\
\> \ttb{defmethod(} method-specifier \ttb{)} \\ \lnk
\> \> method-statement \\ \lnk
\> \ttb{endmethod} \\
\\
method-specifier: \\
\> return-type \ttb{,} method-def \ttb{,} param-list \\
\\
method-def: \\
\> generic-name \\
\> \ttb{(} generic-name \ttb{)} tag-name\opt 	\>\> (around method) \\
\\
method-statement: \\
\> compound-statement 						\>\>\>\> (c99) \\
\> compound-statement-with-contract 			\>\>\> (contract) \\
\\
method-return-statement: \\
\> \ttb{retmethod(} expression\opt\ \ttb{);}\\
\\
method-alias-definition: \\
\> \ttb{defalias(} generic-specifier \ttb{);} \\
\\
alternate-next-method-definition: \\
\> \ttb{defnext(} generic-specifier \ttb{);}\\
\\
next-method-statement: \\
\> \ttb{next\_method(} argument-expression-list \ttb{);}\\
\\
forward-message-statement: \\
\> \ttb{forward\_message(} argument-expression-list \ttb{);}
\end{tabbing}
\vspace{-1mm}}
\end{center}
\caption{Syntax summary of methods.\label{fig:mthgram}}
\end{figure}

\paragraph{Method types}

In order to support fast generic delegation (section~\ref{ssec:fwd}), \Cos must use internally the same function types (\ie same C function signatures) for methods implementation belonging to generics of the same rank:
\begin{COS}
void (*IMP1)(SEL,OBJ,void*,void*);
void (*IMP2)(SEL,OBJ,OBJ,void*,void*);
void (*IMP3)(SEL,OBJ,OBJ,OBJ,void*,void*);
...
\end{COS}
\vspace{-0.2em}\noindent
The first parameter \code{_sel} is the message selector (\ie generic's object) used by the dispatcher, the \code{OBJ}s \code{_}{\em n} are the objects used as selectors (\ie receivers) by the dispatcher, the penultimate parameter \code{_arg} is a pointer to the structure storing the closed arguments of the generic (if any) and the last parameter \code{_ret} is a pointer to the placeholder of the returned value (if any). The responsibilities are shared as follow:

\begin{itemize}
\item The {\em generic functions} are in charge to pack the closed arguments (if any) into the structure pointed by \code{_arg}, to create the placeholder pointed by \code{_ret} for the returned value (if any), to lookup for the method specialization and to invoke its implementation (\ie {\bf\ttfamily IMP}{\em n}) with the prepared arguments \code{_sel}, \code{_}{\em n}, \code{_arg} and \code{_ret}.

\item The {\em methods} are in charge to unpack the closed arguments into local variables and to handle the returned value appropriately.
\end{itemize}

\subsection{Next method}

The \code{next_method} principle borrowed from \Clos\footnote{Namely \closcode{call-next-method}.} is an elegant answer to the problem of superclass(es) methods {\em call} (\ie late binding) in the presence of {\em multi-methods}. The following sample code defines a specialization of the message \code{gincrBy} for the class \code{MilliCounter} which adds thousandths of count to the class \code{Counter}:
\begin{COS}[left]
defmethod(void, gincrBy, MilliCounter, (int)by)
  self->mcnt += by;
  if (self->mcnt >= 1000) {
    defnext(void, gincr, MilliCounter); //*\label{ln:alt}*//
    self->mcnt -= 1000;
    next_method(self); // call gincr(Counter) //*\label{ln:nxt}*//
  }
endmethod
\end{COS}
which is equivalent to the \Objc code:
\begin{OBJC}
- (void) incrBy: (int)by {
  self->mcnt += by;
  if (self->mcnt >= 1000) {
    self->mcnt -= 1000;
    [super incr];
  }
}
\end{OBJC}
Line~\ref{ln:nxt} shows how \Cos \code{next_method} replaces the message sent to \objcode{super} in \Objc. By default, \code{next_method} calls the next method belonging to the same generic (\eg \code{gincrBy}) where {\em next} means the method with the highest specialization less than the current method. But in the example above, the \code{Counter} class has no specialization for \code{gincrBy}. That is why the line~\ref{ln:alt} specifies an {\em alternate next method path}, namely \code{gincr}, to redirect the \code{next_method} call to the appropriate next method. In some cases, it might be safer to test for the existence of the next method before calling it:
\begin{COS}
if (next_method_p) next_method(self);
\end{COS}
It is worth to note that \code{next_method} transfers the returned value (if any) directly from the called next method to the method caller. Nevertheless, the returned value can still be accessed through the {\em lvalue} \code{RETVAL}.

\paragraph{Methods specialization}

Assuming for instance the class inheritance \code{A :> B :> C}, the {\em class precedence list} for the set of all pairs of specialization of \code{A}, \code{B} and \code{C} by {\em decreasing order} will be:
\begin{COS}
(C,C)(C,B)(B,C)(C,A)(B,B)(A,C)(B,A)(A,B)(A,A)
\end{COS}
and the list of {\em all} \code{next_method} {\em paths} are:
\begin{COS}
(C,C)(C,B)(C,A)(B,A)(A,A)
(B,C)(B,B)(B,A)(A,A)
(A,C)(A,B)(A,A)
\end{COS}
The algorithm used by \Cos to build the class precedence list (\ie compute methods rank) has some nice properties: it provides natural asymmetric {\em left-to-right precedence} and it is {\em non-ambiguous}, {\em monotonic} and {\em totally ordered} \cite{c3}.

\paragraph{Around methods}

Around methods borrowed from \Clos provide an elegant mechanism to enclose the behavior of some {\em primary method} by an arbitrary number of around methods. Around methods are always {\em more specialized} than their primary method but have an undefined precedence:
\begin{COS}
defmethod(void, gdoIt, A, A)
endmethod

defmethod(void, gdoIt, B, A)
  next_method(self1, self2);  // call gdoIt(A,A)
endmethod

defmethod(void, (gdoIt), B, A) // around method
  next_method(self1, self2); // call gdoIt(B,A)
endmethod

defmethod(void, gdoIt, B, B)
  next_method(self1, self2); // call (gdoIt)(B,A)
endmethod
\end{COS}

\subsection{Delegation\label{ssec:fwd}}

Message forwarding is a major feature of \Cos which was developed from the beginning with {\em fast generic delegation} in mind as already mentioned in the previous section.

\paragraph{Unrecognized message}

Message dispatch performs runtime look\-up to search for method specializations. If no specialization is found, the message \code{gunrecognizedMessage}{\em n} is {\sc substituted} and {\em sent with the same arguments as the original sending, including the selector}. Hence these messages can be overridden to support the delegation or some adaptive behaviors. The default behavior of \code{gunrecognizedMessage}{\em n} is to throw the exception \code{ExBadMessage}.

\paragraph{Forwarding message}

Message forwarding has been borrowed from \Objc and extended to multi-methods. The sample code below shows a common usage of message forwarding to protect objects against invalid messages:
\begin{COS}[left]
defmethod(void, gunrecognizedMessage1, MyProxy)
  if(gundertstandMessage1(self->obj,_sel)==True)
    forward_message(self->obj); // delegate
endmethod
\end{COS}
which can be translated line-by-line into \Objc by:
\begin{OBJC}[left]
- (retval_t) forward:(SEL)sel :(arglist_t)args {
  if ([self->obj respondsTo: sel] == YES)
    return [self->obj performv:sel :args];
}
\end{OBJC}
Here, \code{forward_message} propagates all the arguments, including the hidden parameters \code{_sel}, \code{_arg} and \code{_ret}, to a different receiver. As for \code{next_method}, \code{forward_message} transfers the returned value directly to the method caller and can be accessed through \code{RETVAL} in the same way.

\paragraph{Fast delegation}

Since all methods belonging to generics with equal rank have the same C function signature and fall into the same lookup cache, it is safe to cache the message \code{gunrecognizedMessage}{\em n} in place of the {\em unrecognized message}. Hence, the next sending of the latter will result in a cache hit.

\vspace{1em}\noindent
\fbox{\parbox{0.98\columnwidth}{{\em
This substitution allows the delegation to be as fast as message dispatch, seemingly a unique feature.}}}
\vspace{0ex}

\paragraph{Intercession of forwarded messages}

Since the closed arguments of the generic's {\em param-list} are managed by a C structure, it is possible to access each argument separately. In order to do this, \Cos provides introspective information on generics (\ie metadata on types and signatures) which allows identifying and retrieving the arguments and the returned value efficiently. But this kind of needs should be exceptional and is beyond the scope of this paper.





\subsection{Contracts\label{ssec:ctr}}

To quote Bertrand Meyer \cite{mey97}, the key concept of Design by Contract is ``{\em viewing the relationship between a class and its clients as a formal agreement, expressing each party's rights and obligations}''. Most languages that support Design by Contract provide
two types of statements to express the obligations of the caller and the callee: {\em preconditions} and {\em postconditions}. The caller must meet all preconditions of the message sent, and the callee (the method) must meet its own postconditions --- the failure of either party leads to a bug in the software. In that way, {\em Design by Contract} (\ie developer point of view) is the complementary tool of {\em Unit Testing} \cite{ken02} (\ie user point of view) and they both enhance the mutual confidence between developers and users, help to better identify the responsibilities and improve the design of interfaces.

\begin{figure}\hr
\begin{center}
{\em
\begin{tabbing}
xxx\= xxx\= \hspace*{19em}\= \kill
compound-statement-with-contract: \\
\> declaration-without-initializer 	\>\> (c99) \\ \lnk
\> pre-statement\opt post-statement\opt body-statement \\
\\
pre-statement: \\
\> \ttb{PRE} statement \\
\\
post-statement: \\
\> \ttb{POST} statement \\
\\
body-statement: \\
\> \ttb{BODY} statement \\
\\
test-assert-statement: \\
\> \ttb{test\_assert(} bool-expr \ttb{)} \\
\> \ttb{test\_assert(} bool-expr \ttb{,} cstr \ttb{)} \\
\> \ttb{test\_assert(} bool-expr \ttb{,} func \ttb{,} file \ttb{,} line \ttb{)} \\
\> \ttb{test\_assert(} bool-expr \ttb{,} cstr \ttb{,} func \ttb{,} file \ttb{,} line \ttb{)} \\
\\
test-invariant-statement: \\
\> \ttb{test\_invariant(} object-expr \ttb{)} \\
\> \ttb{test\_invariant(} object-expr \ttb{,} func \ttb{,} file \ttb{,} line \ttb{)}
\end{tabbing}}
\end{center}
\caption{Syntax summary of contracts.\label{fig:ctrgram}}
\end{figure}

To illustrate how contracts work in \Cos with the syntax summarized in figure~\ref{fig:ctrgram}, we can rewrite the method \code{gincr}:
\begin{COS}
defmethod(void, gincr, Counter)
  int  old_cnt; // no initializer!
  PRE  old_cnt = self->cnt;
  POST test_assert(self->cnt < old_cnt);
  BODY self->cnt++;
endmethod
\end{COS}
The \code{POST} statement \code{test_assert} checks for counter overflow {\em after} the execution of the \code{BODY} statement and throws an \code{ExBadAssert} exception on failure, breaking the contract. The variable \code{old_val} initialized in the \code{PRE} statement {\em before} the execution of the \code{BODY} statement, plays the same role as the \eiffelcode{old} feature in \Eiffel. As well \code{gincrBy} can be improved:

\vspace{2mm}
\begin{COS}
defmethod(void, gincrBy, MilliCounter, (int)by)
  PRE test_assert(by >= 0 && by < 1000,
                    ''millicount out or range'');
  BODY // same code as before
endmethod
\end{COS}
The \code{PRE} statement ensures that the {\em incoming} \code{by} is within the expected range and the \code{next_method} call in the \code{BODY} statement ensures that the contract of \code{gincr} is also fulfilled.

\paragraph{Assertions and tests}

In order to ease the writing of contracts and unit tests, \Cos provides two standard tests:
\begin{itemize}
\item \code{test_assert(expr[,str][,func,file,line])} is a replacement for the standard \code{assert} and raises an \code{ExBadAssert} exception on failure. The (optional) parameters {\em str, func, file} and {\em line} are transfered to \code{THROW} for debugging.


\item \code{test_invariant(obj[,func,file,line])} checks for the {\em class invariants} of objects. It can only be used inside methods and is automatically invoked on each receiver if the invariant contract level is active. The (optional) parameters {\em func, file} and {\em line} are transfered to \code{ginvariant}.
\end{itemize}


\paragraph{Class invariants}

The \code{test_invariant} assertion relies on the message \code{ginvariant} which must be specialized for \code{MilliCounter} to be effective in the previous example:
\begin{COS}
defmethod(OBJ, ginvariant, MilliCounter, 
//*\hfill*// (STR)func, (STR)file, (int)line)
  next_method(self); // check Counter invariant
  int mcnt = self->mcnt;
  test_assert(mcnt >= 0 && mcnt < 1000,
  ''millicount out of range'', func, file, line);
endmethod
\end{COS}
Here, \code{test_assert} propagates the location of the calling \code{test_invariant} to improve bug tracking. 

\paragraph{Contracts and inheritance}

In the design of \Eiffel, Ber\-trand Meyer recommends to evaluate inherited contracts as a {\em disjunction} of the preconditions and as a {\em conjunction} of the postconditions. But \cite{find01} demonstrates that \Eiffel-style contracts may introduce behavioral inconsistencies with inheritance, thus \Cos prefers to treat both pre and post conditions as conjunctions. This is also the only known solution compatible with multi-methods where subtyping is superseded by the {\em class precedence list}.

\paragraph{Contracts levels}

The level of contracts can be set by defining the macro \code{COS_CONTRACT} to one of the levels:
\begin{itemize}
\item \code{NO} disable contracts (not recommended).

\item \code{COS_CONTRACT_PRE} enables \code{PRE} sections. This is the recommended level for {\em production phases} (default level).

\item \code{COS_CONTRACT_POST} enables \code{PRE} and \code{POST} sections. This is the usual level during the {\em development phases}.

\item \code{COS_CONTRACT_ALL} enables \code{PRE} and \code{POST} sections as well as \code{test_invariant} statements. This is the highest level usually set during {\em debugging phases}.
\end{itemize}

\begin{figure}\hr
\begin{center}
{\em
\begin{tabbing}
xxx\= xxx\= \hspace*{17em}\= \hspace*{2em} \= \kill
property-declaration: \\
\> \ttb{useproperty(} property-decl-list \ttb{);} \\
\\
property-decl-list: \\
\> property-decl \\
\> property-decl-list \ttb{,} property-decl \\
\\
property-decl: \\
\> property-name \\
\> \ttb{(} property-name \ttb{)} local-name \\
\\
property-definition: \\
\> \ttb{defproperty(} property-def \ttb{);} \\
\\
property-def: \\
\> property-name \\
\> \ttb{(} super-property-name \ttb{)} property-name \\
\\
class-property-definition: \\
\> \ttb{defproperty(} class-property-def \ttb{);} \\
\\
class-property-def: \\
\> class-name \ttb{,} property-attr \\
\> class-name \ttb{,} property-attr \ttb{,} get-func\opt \\
\> class-name \ttb{,} property-attr \ttb{,} get-func\opt \ttb{,} put-func\opt \\
\\
property-attr: \\
\> property-name \\
\> \ttb{(} object-attribute\opt \ttb{)} property-name \\
\\
\{property, super-property\}-name, object-attribute: \\
\> identifier					\>\>\> (c99)
\end{tabbing}
}
\end{center}
\caption{Syntax summary of properties.\label{fig:prpgram}}
\end{figure}

\subsection{Properties\label{ssec:prp}}

Property declaration is a useful programming concept which allows, amongst others, to manage the access of object attributes, to use objects as associative arrays or to make objects persistent. Figure~\ref{fig:prpgram} summarizes the syntax of properties in \Cos which are just syntactic sugar on top of the definition of class-objects and the specialization of the accessors \code{ggetAt} and \code{gputAt} already mentioned in section~\ref{ssec:gen}.

\paragraph{Property definition}

Properties in \Cos are defined conventionally with lowercase names:
\begin{COS}
defproperty( name );
defproperty( size );
defproperty( class );
defproperty( value );
\end{COS}
For example, the last property definition is equivalent to:
\begin{COS}
defclass(P_value, Property)
endclass
\end{COS}
Most notably, properties are class-objects deriving from the class Property (fig. \ref{fig:clstree}) with lowercase names prefixed by \code{P_}.

\paragraph{Class properties}

Once properties have been defined, it is possible to define some class-properties:
\begin{COS}
defproperty(Counter, (cnt)value, int2OBJ, gint);
defproperty(Counter, ()class, gclass);
\end{COS}
\vspace{-1.2mm}with:\vspace{-0.9mm}
\begin{COS}
OBJ int2OBJ(int val) { // cannot be a method
  return gautoDelete(aInt(val));
}
\end{COS}
The \code{value} property is associated with the \code{cnt} attribute with read-write semantic and uses user-defined boxing (\code{int2OBJ}) and unboxing (\code{gint}). The \code{class} property is associated with the entire object (omitted attribute) with read-only semantic and uses the inherited message \code{gclass} to retrieve it.

Sometimes the abstraction or the complexity of the properties require handwritten methods. For instance:
\begin{COS}
defmethod(OBJ, ggetAt, Person, mP_name)
  retmethod(gcat(self->fstname, self->lstname));
endmethod
\end{COS}
is equivalent to, assuming \code{gname(Person)} is doing the \code{gcat}:
\begin{COS}
defproperty(Person, ()name, gname);
\end{COS}

\paragraph{Using properties}

The example below displays the \code{name} property of an object (or raise the exception \code{ExBadMessage}): 
\begin{COS}
void print_name(OBJ obj) {
  useproperty(name);
  gprint(ggetAt(obj, name));
}
\end{COS}

\section{Exceptions\label{sec:ex}}

Exceptions are non-local errors which ease the writing of interfaces since they allow {\em solving the problems where the solutions exist}. To state it differently, if an {\em exceptional} condition is detected, the callee needs to return an error and let the caller take over. Applying recursively this behavior requires a lot of boilerplate code on the callers side to check returned status. Exceptions let the callers choose to either ignore {\em thrown} errors or to {\em catch} them and take over.

Implementing an exception mechanism in C on top of the standard \code{setjmp} and \code{longjmp} is not new. But it is uncommon to see a framework written in C which provides the full {\em try-catch-finally} statements (figure~\ref{fig:exgram}) with the same semantic as in other object-oriented programming languages (\eg\ \Java, \Csharp). The \code{CATCH} declaration relies on the message \code{gisKindOf} to identify the thrown exception, what implies that the order of \code{CATCH} definitions matters, as usual.

\begin{figure}\hr
\begin{center}
{\em
\begin{tabbing}
xxx\= xxx\= \hspace*{19em}\= \kill
try-statement: \\
\> \ttb{TRY} \\ \lnk
\> \> statement \\ \lnk
\> \> catch-statement-list\opt \\ \lnk
\> \> finally-statement\opt \\ \lnk
\> \ttb{ENTRY} \\
\\
catch-statement-list: \\
\> catch-statement \\
\> catch-statement-list catch-statement \\
\\
catch-statement: \\
\> \ttb{CATCH(} class-name \ttb{,} exception-name\opt\ \ttb{)} statement \\
\> \ttb{CATCH\_ANY(} exception-name\opt\ \ttb{)} statement \\
\\
finally-statement: \\
\> \ttb{FINALLY} statement \\
\\
throw-statement: \\
\> \ttb{THROW(} object-expr \ttb{);} \\
\> \ttb{THROW(} object-expr \ttb{,} func \ttb{,} file \ttb{,} line \ttb{);} \\
\> \ttb{RETHROW();} \\
\\
exception-name: \\
\> identifier		\>\> (c99)
\end{tabbing}
}
\end{center}
\caption{Syntax summary of exceptions.\label{fig:exgram}}
\end{figure}

The sample program hereafter gives an overview of exceptions in \Cos:
\begin{COS}[left]
int main(void) {
  useclass(String, ExBadAssert, mExBadAlloc);
  STR s1 = 0;
  OBJ s2 = Nil;
  
  TRY
    s1 = strdup(''str1'');
    s2 = gnewWithStr(String, ''str2'');
    test_assert(0, ''throw ExBadAssert'');

  CATCH(ExBadAssert, ex)
    printf(''assertion %s failed (%s,%d)\n'',
            gstr(ex), ex_file, ex_line);
    gdelete(ex);
  CATCH(mExBadAlloc, ex) // catch class ExBadAlloc//*\label{ln:alloc}*//
    printf(''out of memory (%s,%d)\n'',
            ex_file, ex_line);
    gdelete(ex);
  CATCH_ANY(ex)
    printf(''unexpected exception %s (%s,%d)\n'',
            gstr(ex), ex_file, ex_line);
    gdelete(ex);   
  FINALLY // always executed //*\label{ln:finally}*//
    free(s1);
    gdelete(s2);
  ENDTRY
}
\end{COS}
The code above shows some typical usages:
\begin{itemize}
\item Line~\ref{ln:alloc} catches the {\em class} \code{ExBadAlloc} which is thrown when a memory allocation failure occurs. Throwing an {\em instance} of the class in a such context would not be safe.

\item Line~\ref{ln:finally} destroys the two strings whatever happened. Their initial states have been set to be neutral for these operations in case of failure.
\end{itemize}
\Cos allows throwing {\em any kind of object} but it provides also a hierarchy of exceptions deriving from \code{Exception}:
{\small\tt Ex\-Bad\-Alloc},
{\small\tt Ex\-Bad\-Arity},
{\small\tt Ex\-Bad\-Assert},
{\small\tt Ex\-Bad\-Cast},
{\small\tt Ex\-Bad\-Domain},
{\small\tt Ex\-Bad\-Format},
{\small\tt Ex\-Bad\-Mes\-sage},
{\small\tt Ex\-Bad\-Pro\-perty},
{\small\tt Ex\-Bad\-Range},
{\small\tt Ex\-Bad\-Size},
{\small\tt Ex\-Bad\-Type},
{\small\tt Ex\-Bad\-Value},
{\small\tt Ex\-Not\-Found},
{\small\tt Ex\-Not\-Imple\-men\-ted},
{\small\tt Ex\-Not\-Sup\-por\-ted},
{\small\tt Ex\-Errno},
and
{\small\tt Ex\-Signal}.
Among these exceptions, \code{ExErrno} and \code{ExSignal} are special cases used respectively to convert standard {\em errors} (\ie \code{test_errno()}) and registered {\em signals} into exceptions.

\section{Performance\label{sec:perf}}

\begin{figure}\hr
\begin{center}
\vspace{-2mm}
\includegraphics[width=0.9\columnwidth,keepaspectratio=true]{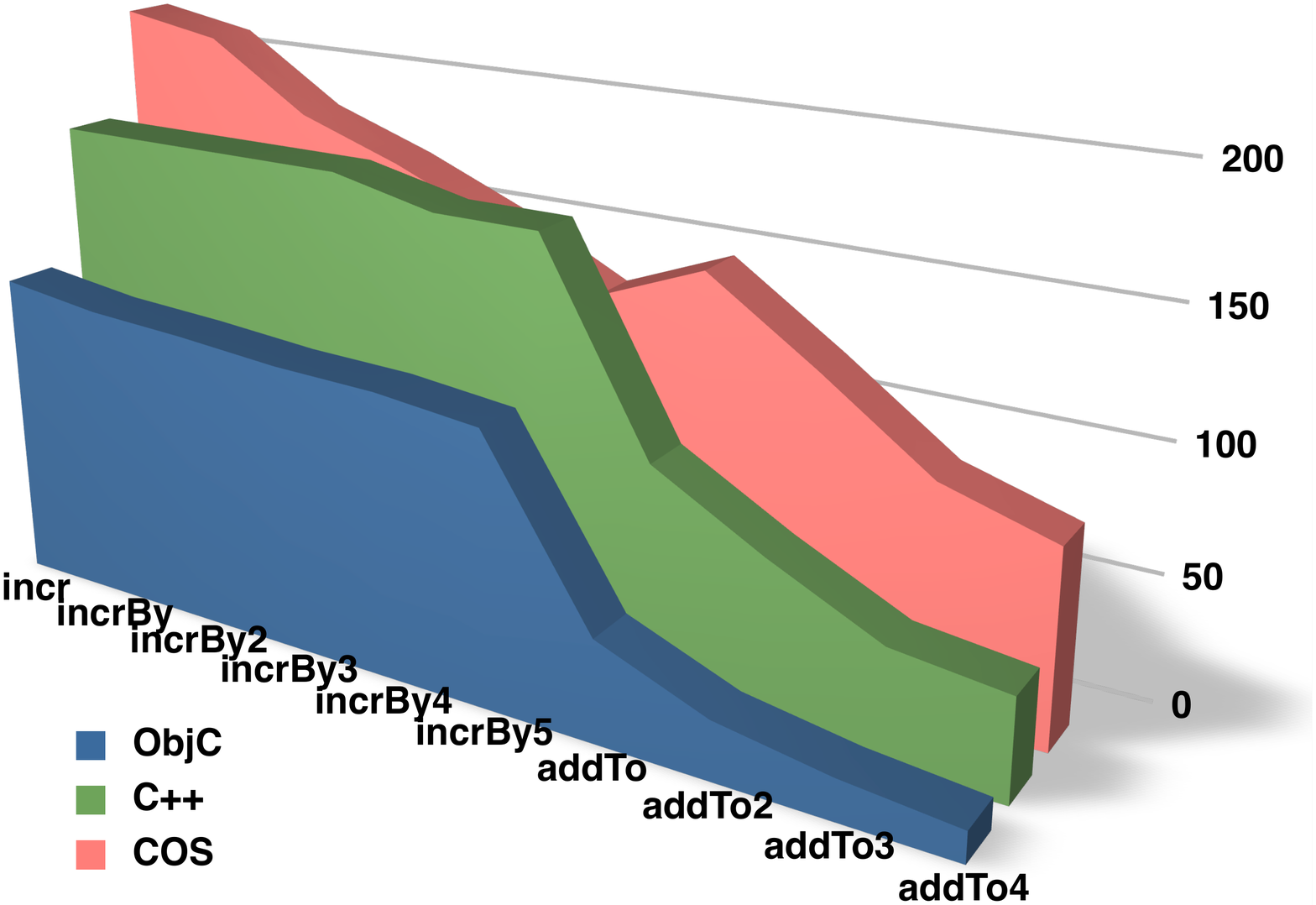}
\vspace{-2.5mm}
\end{center}
\caption{Performances summary in $10^6$ calls/second.\label{fig:perf}}
\end{figure}

In order to evaluate the efficiency of \Cos, small test suites\footnote{These testsuites can be browsed on sf.net in the module {\tt CosBase/tests}.} have been written to stress the message dispatcher in various conditions. The test results summarized in table~\ref{tbl:perf} and figure~\ref{fig:perf} have been performed on an Intel DualCore2\texttrademark{} T9300 \Cpu 2.5 Ghz with Linux Ubuntu 64-bit and the compiler \Gcc 4.3 to compile the tests written in the three languages. The timings have been measured with \code{clock()} and averaged over $10$ loops of $2 \cdot 10^8$ iterations each. The {\em Param.} column indicates the number of parameters of the message split by {\em selectors} (open types) and {\em arguments} (closed types). The other columns represent the performances in million of invocations sustained per second for respectively \Cpp virtual member functions, \Objc messages and \Cos messages. The tests stress the dispatcher with messages already described in this paper: \code{incr} increments a counter, \code{incrBy\{2..5\}}\opt accept from 1 to 5 extra {\em closed} parameters (to stress the construction of \code{_arg}) and \code{addTo\{2..4\}}\opt add from 2 to 5 \code{Counter}s together (to stress multiple dispatch). Multiple dispatch has been implemented with the Visitor design pattern in \Cpp and \Objc.

\begin{table}
\begin{center}
{\small
\begin{tabular}{l|c|ccc} \hline
\multicolumn{1}{l}{Tests}& \multicolumn{1}{l}{Param.}
                                            & \Cpp & {\sc ObjC} & \Cos \\ \hline
\multicolumn{5}{c}{{\em single dispatch}} \\
\code{counter incr}          & $1+0$        & 176  & 122        & 218  \\
\code{counter incrBy}        & $1+1$        & 176  & 117        & 211  \\
\code{counter incrBy2}       & $1+2$        & 176  & 115        & 185  \\
\code{counter incrBy3}       & $1+3$        & 176  & 112        & 171  \\
\code{counter incrBy4}       & $1+4$        & 167  & 111        & 154  \\
\code{counter incrBy5}       & $1+5$        & 167  & 107        & 133  \\ \hline
\multicolumn{5}{c}{{\em multiple dispatch}} \\
\code{counter addTo}         & $2+0$        & 90   & 40         & 150  \\
\code{counter addTo2}        & $3+0$        & 66   & 23         & 121  \\
\code{counter addTo3}        & $4+0$        & 45   & 16         & 90   \\
\code{counter addTo4}        & $5+0$        & 40   & 12         & 77   \\
\end{tabular}}
\end{center}
\caption{Performances summary in $10^6$ calls/second.\label{tbl:perf}}
\end{table}

Concerning the performance of {\em single dispatch}, \Cos shows a good efficiency since it runs in average at about the same speed as \Cpp and about $\times 1.6$ faster than \Objc. On one hand, \Cos efficiency decreases faster than \Cpp effciency because it passes more hidden arguments (\ie \code{_sel} and \code{_arg}) and uses more registers to compute the dispatch. On the other hand, \Cpp shows some difficulties to manage efficiently multiple inheritance of abstract classes.
Concerning the performance of {\em multiple dispatch}, \Cos outperforms \Cpp and \Objc by factors $\times 1.9$ and $\times 5.3$.
Concerning the performance of {\em message forwarding}, we have seen that by design, it runs at the full speed of message dispatch in \Cos. Rough measurements of \Objc message forwarding (linear lookup) shows that \Cos performs from $\times 50$ to $\times 100$ faster, depending on the tested classes.

\paragraph{Multi-threading}

The same performance tests have been run with \Posix multi-threads enabled. When the Thread-Local-Storage mechanism is available (Linux), no significant impact on performance has been observed ($<$1\%). When the architecture supports only \Posix Thread-Specific-Key (Mac OS X), the performance is lowered by a factor $\times 1.6$ and becomes clearly the bottleneck of the dispatcher.

\paragraph{Object creation}

Like other languages with semantic by reference, \Cos loads heavily the C memory allocator (\eg \code{malloc}) which is not very fast. If the allocator is identified as the bottleneck, it can be replaced with optimized pools by overriding \code{galloc} or by faster external allocators (\eg Google \code{tcmalloc}). \Cos also takes care of automatic objects which can be used to speed up the creation of local objects.

\paragraph{Other aspects}

Other features of \Cos do not involve such heavy machinery as in message dispatch or object creation. There\-by, they all run at full speed of C. Contracts run at the speed of the user tests since the execution path is known at compile time and flattened by the compiler optimizer. Empty try-blocks run at the speed of \code{setjmp} which is a well known bottleneck. Finally \code{next_method} runs at $70\%$ of the speed of an indirect function call (\ie late binding) because it also has to pack the closed arguments into the generic's \code{_arg} structure.

\section{Component-Oriented Design Patterns\label{sec:pat}}

This overview of \Cos shows that the principles stated in the introduction are already well fulfilled. So far:

\begin{itemize}
\item The {\em simplicity} can be assumed from the fact that the entire language can be described within few pages, including the grammar, some implementation details and few examples and comparisons with other languages.

\item The {\em extensibility} comes from the nature of the object model which allows extending (methods bound to generics), wrapping (around methods) or renaming (method aliases) behaviors with a simple user-friendly syntax. Besides, encapsulation, polymorphism, low coupling (messages) and contracts are also assets for this principle.

\item The {\em reusability} comes from the key concepts of \Cos which enhance generic design: polymorphism, collaboration (multi-methods) and composition (delegation).

\item The {\em efficiency} measurement shows that key concepts perform well compared to other mainstream languages.

\item The {\em portability} comes from its nature: a C89 library.
\end{itemize}

It is widely acknowledged that dynamic programming languages simplify significantly the implementation of classical design patterns \cite{gof95} when they don't supersede them by more powerful dynamic patterns \cite{cocoa07,nor96,sul02}. This section focuses on how to use \Cos features to simplify design patterns or to turn them into reusable components, where the definition of {\em componentization} is borrowed from \cite{rege99,meyer06-1}:

\vspace{1em}\noindent
\fbox{\parbox{0.98\columnwidth}{{\em
Component Orientation = Encapsulation + Polymorphism \\
\mbox{} \hfill + Late Binding + Multi-Dispatch + Delegation.}}}
\vspace{1ex}


\subsection{Simple Patterns\label{ssec:spat}}

\paragraph{Creational Patterns}

It is a well known fact that these patterns vanish in languages supporting generic types and introspection. We have already seen \code{gnew} (p. \pageref{par:new}), here is more:
\begin{COS}
OBJ gnewWithStr(OBJ cls, STR str) {
  return ginitWithStr(galloc(cls), str);
}

OBJ gclone(OBJ obj) {
  return ginitWith(galloc(gclass(obj)), obj);
}
\end{COS}
The Builder pattern is a nice application of property metaclasses to turn it into the so called Class Cluster pattern:
\begin{COS}
defmethod(OBJ, galloc, pmString)
  retmethod(_1); // lazy, delegate the task to initializers
endmethod

defmethod(OBJ, ginitWithStr, pmString, (STR)str)
  OBJ lit_str = galloc(StringLiteral);
  retmethod( ginitWithStr(lit_str, str) );
endmethod
\end{COS}
This example shows how to delegate the object build to the initializer which in turn allocates the appropriate object according to its arguments, an impossible task for the allocator. The allocation of the \code{StringLiteral} uses the standard allocator inherited from \code{Object}, even if it derives from the class \code{String} (thanks to property metaclass). Now, the code:
\begin{COS}
OBJ str = gnewWithStr(String,''literal string'');
\end{COS}
will silently return an instance of \code{StringLiteral}. This is the cornerstone of Class Clusters where the front class (\eg \code{String}) delegates to private subclasses (\eg \code{StringLiteral}) the responsibility to build the object. It is worth to note that each \code{pmString} specialization needed to handle new subclass is provided by the subclass itself (thanks to the open class model), which makes the Builder pattern truly extensible. Most complex or multi-purpose classes of \Cos are designed as class clusters (\eg \code{Array}, \code{String}, \code{Functor}, \code{Stream}).

\paragraph{Garbage Collector}

This exercise will show how to simplify memory management in \Cos with only few lines of code. We start by wrapping the default object allocator such that it {\em always} auto-releases the allocated objects:
\begin{COS}
defmethod(OBJ, (galloc), mObject) // around method
  next_method(self); // allocate
  gautoRelease(RETVAL); // auto-release
endmethod
\end{COS}
Then we neutralize (auto-)delete and reinforce retain:
\begin{COS}
defmethod(void, (gdelete), Object)
endmethod

defmethod(OBJ, (gautoDelete), Object)
  BOOL is_auto = self->rc == COS_RC_AUTO;
  retmethod( is_auto ? gclone(_1) : _1 );
endmethod

defmethod(void, (gretain), Object)
  next_method(self);
  if (self->rc == COS_RC_AUTO)
    RETVAL = gretain(RETVAL); // once more for auto
endmethod
\end{COS}
Now, the following code:
\begin{COS}
OBJ pool = gnew(AutoRelease);
for(int i = 0; i < 1000; i++)
  OBJ obj = gnewWithStr(String, ''string'');
gdelete(pool); // pool specialization, collect the strings
\end{COS}
does not create any memory leak, there is no longer the need to delete or auto-delete your objects. For the first runs, you can rely on the default auto-release pool managed by \Cos. Then a profiler of used memory will show the appropriate locations where intermediate auto-release pools should be added to trigger the collects and limit the memory usage.

\paragraph{Key-Value-Coding}

We have already seen that properties allow accessing to object attributes, but to implement KVC, we need to translate strings (key) into properties (noun):
\begin{COS}
defmethod(OBJ, ggetAt, Object, String)
  OBJ prp = cos_property_getWithStr(self2->str);
  if (!prp) THROW(gnewWith(ExBadProperty,_2));
  retmethod( ggetAt(_1, prp) );
endmethod

defmethod(void, gputAt, Object, String, Object)
  OBJ prp = cos_property_getWithStr(self2->str);
  if (!prp) THROW(gnewWith(ExBadProperty,_2));
  gputAt(_1, prp, _3);
endmethod
\end{COS}
where \code{cos_property_getWithStr} is an optimized version of \code{cos_class_getWithStr} for properties from the API of \Cos, which also provides \code{cos_class_\{read,write\}Properties} to retrieve all the properties of a class (and its superclasses).

\paragraph{Key-Value-Observing}

Adding access notifications of properties is the next step after KVC, using around methods:
\begin{COS}
defmethod(OBJ, ggetAt, (Person), mP_name)
  useclass(After, Before);
  OBJ context = aMthCall(_mth,_1,_2,_arg,_ret);
  gnotify(center, context, Before);
  next_method(self1, self2); // get property
  gnotify(center, context, After);
endmethod
\end{COS}
where \code{_mth} is the object representing the method itself. This example assumes that observers and objects observed have already been registered to some notification \code{center} as commonly done in the Key-Value-Observing pattern.



\subsection{Proxies and Decorators\label{ssec:ppat}}

\paragraph{Proxy}

Almost all proxies in \Cos derive from the class \code{Proxy} which handles some aspects of this kind of class:
\begin{COS}
defclass(Proxy);
  OBJ obj; // delegate
endclass

#define gum1 gunrecognizedMessage1 // shortcut
#define gum2 gunrecognizedMessage2 // shortcut

defmethod(void, gum2, Proxy, Object)
  forward_message(self1->obj, _2);
  check_ret(_sel, _ret, self1);
endmethod

defmethod(void, gum2, Object, Proxy)
  forward_message(_1, self2->obj);
  check_ret(_sel, _ret, self2);
endmethod

// ... other rank specializations
\end{COS}
where the small function \code{check_ret} takes care to return the proxy when the forwarded message returns the delegate \code{obj}. 

\paragraph{Tracer}

For the purpose of debugging, \Cos provides the simple proxy \code{Tracer} to trace the life of an object:
\begin{COS}
defclass(Tracer,Proxy) // usage: gnewWith(Tracer, obj);
endclass

defmethod(void, gum2, Tracer, Object)
  trace_msg2(_sel, self1->Proxy.obj, _2);
  next_method(self1, self2); // forward message
endmethod

defmethod(void, gum2, Object, Tracer)
  trace_msg2(_sel, _1, self2->Proxy.obj);
  next_method(self1, self2); // forward message
endmethod

// ... other rank specializations
\end{COS}
where \code{trace_msg2} prints useful information on the console.

\paragraph{Locker}

The locker is a proxy which {\em avoids} synchronization deadlock on shared objects that can be encountered in programming languages supporting only single-dispatch \cite{clos89}: 

\begin{COS}
defclass(Locker,Proxy) // usage: gnewWith(Locker, obj);
  pthread_mutex_t mutex;
endclass

defmethod(void, gum1, Locker) // like smart pointers
  lock(self); // lock the mutex
  next_method(self); // forward the message
  unlock(self); // unlock the mutex
endmethod

defmethod(void, gum2, Locker, Locker)
  sorted_lock2(self1, self2); // lock in order
  next_method(self1, self2); // forward the message
  sorted_unlock2(self1, self2); // unlock in order
endmethod

// ... other rank specializations (total of 63-6 combinations)
\end{COS}
For the sake of efficiency, higher ranks use {\em sorting networks}.

\paragraph{Multiple Inheritance}

The first version of \Cos was natively implementing multiple inheritance using the C3 algorithm \cite{c3} to compute the {\em class precedence list} on the way of \Dylan, \Python and \PerlSix. But it was quickly considered as too complex for the end-user and incidental as far as fast generic delegation could be achieved. Indeed, multiple inheritance can be simulated by {\em composition} and {\em delegation} with an efficiency close to native support\footnote{\Objc delegation is too slow to simulate multiple inheritance.}:
\begin{COS}
defclass(IOStream, OutStream)
  OBJ in_stream;
endclass

defmethod(void, gum1, IOStream)
  forward_message(self->in_stream);
endmethod
\end{COS}
Now, messages of rank one not understood by the \code{IOStream}s (\eg \code{gget}, \code{gread}) will be forwarded to their \code{InStream}.

\paragraph{Distributed Objects} Without going into the details, we can mention that \Cos already implements all the key concepts required to develop a distributed object system on the model of {\sc Objec\-tive-C} and \Cocoa. A challenge for the future.






\subsection{Closures and Expressions\label{ssec:clo}}

\Cos provides the family of \code{geval}{\em n} messages (equivalent to \CL \closcode{funcall}) and the class cluster \code{Functor} to support the mechanism of closures and more generally lazy expressions and high order messages.
The objects representing the context of the closure (\ie the free variables) are passed to the \code{Functor} constructor which handles partial evaluation and build expressions.
The next example shows another way to create a counter in \Perl using a closure:
\begin{PERL}[left]
sub counter {
  my($val) = shift; # seed
  $cnt = sub { # incr
    return $val++;
  };
  return $cnt; # return the closure
}

$cnt = counter(0);
for($i=0; $i<25000000; $i++)
  &$cnt();
\end{PERL}
which can be translated into \Cos as:
\begin{COS}[left]
OBJ counter(int seed) {
  OBJ fun = aFunctor(gincr, aCounter(seed));//*\label{ln:clo}*//
  return gautoDelete(fun); //*\label{ln:fun}*//
}

int main(void) {
  OBJ cnt = counter(0);
  for(int i=0; i<25000000; i++)
    geval(cnt);
}
\end{COS}
The line~\ref{ln:clo} creates the closure using the automatic constructor \code{aFunctor} which takes the generic function \code{gincr} and {\em deduces} its arity (here 1) from the remaining parameters, namely the seed {\em boxed} in the counter. Line~\ref{ln:fun}, the message \code{gautoDelete} extends the lifespan of both the functor and the counter to the dynamic scope. As one can see, \Cos achieves the same task as \Perl with about the same amount of code but runs more than $\times 15$ {\em faster}. The example below:
\begin{COS}
fun = gaddTo(Var,y) // works only with messages
fun = aFunctor(gaddTo,Var,y) // works with functions
\end{COS}
create a closure with {\em arity 2} where \code{Var} specifies an argument placeholder and \code{y} an argument object. The message \code{geval1(fun,x)} is then equivalent to \code{gaddTo(x,y)}. 

The following example shows a more advanced example involving lazy expressions and indexed placeholders:
\newpage
\begin{COS}
// return //*$f' = ( f(x + dx) - f(x) ) / dx$*//
OBJ ggradient(OBJ f) {
  OBJ x = aVar(0); // placeholder #1
  OBJ dx = aVar(1); // placeholder #2
  OBJ f_x = geval1(f, x); // lazy expression
  OBJ f_xpdx = geval1(f, gadd(x, dx)); // idem
  return gdiv(gsub(f_xpdx, f_x), dx); // idem
}
//* \vspace{-1ex} *//
// return //*$f'(x)|_{dx}$*//
OBJ gderivative(OBJ f, OBJ dx) {
  return geval2(ggradient(f), Var, dx);
}
\end{COS}
Now we can map this function to a \code{bag} of values (strict evaluation) or expressions (lazy evaluation) indifferently:
\begin{COS}
OBJ fun = gderivative(f, dx);
OBJ new_bag = gmap(fun, bag);
\end{COS}

\paragraph{High Order Messages}

The principle behind {\Hom}s is to construct on-the-fly an expression from their arguments composition  \cite{weih05} --- a technique known for more than a decade in \Cpp meta-programming. Once the expression is completed, the last built object evaluates the expression and returns the result. While \Cpp meta-expressions rely strongly on static types (\ie traits) and templates to build and simplify the expressions, {\Hom}s rely on the delegation mechanism:
\begin{itemize}
\item With fast generic delegation, no need to cache the message in the \Hom objects as in the aforementioned paper.

\item With multi-methods, no need to provide multiple {\Hom}s for similar tasks (\ie \code{gfilter}, \code{gselect} and \code{gcollect}).

\item With lazy expression, no need to construct complex {\em meta-expressions} or to reorder {\em compositions}.
\end{itemize}
{\Hom}s are an important tool for modern framework design since they play the role of weavers of {\em cross-cutting concerns} otherwise solved by foreign technologies based on subject-oriented \cite{har93} and aspect-oriented programming \cite{kic97}. Likewise, the availability of {\Hom}s simplify drastically the implementation of interpreters reflecting the classes and methods of the underlying language like in \Fscript \cite{mou06}.






\section{Conclusion\label{sec:end}}

Some frameworks, with the help of external preprocessors or compilers, propose extensions to the C programming language, but none of them provide a set of features as consistent, complete and efficient as \Cos. Besides, even if the features of \Cos are not new, I am not aware of a single programming language which provides the same set of features with the same {\em simplicity, portability} and {\em efficiency}. Finally, very few programming languages target all the principles stated in the introduction. In particular, modern type systems which try to improve code flexibility and reusability to ease the design of generic components, tend to be lazy (static duck typing), overcomplicated (C++ {\sc adl}) and counter productive for average or occasional developers.

\subsection{Related work}

\paragraph{Ooc}

This old framework uses the \code{ooc} preprocessor written in \Awk \cite{ooc94} to provide a basic object-oriented layer. It relies on \code{void} pointers and requires much defensive programming to ensure correct use of objects. Besides, it gives full control over inheritance 
like in prototype-based languages.

\paragraph{Dynace}

This framework includes the \code{dpp} preprocessor and a large library of classes \cite{dyn06}. Dynace features are equivalent to those of \Objc except that it supports multiple inheritance. However, Dynace message dispatch is about $\times 3$ {\em slower} than \Cos even with {\em jumpTo} assembly code enabled and accessing object attributes is a bit awkward and relies on fancy macros (\eg \code{accessIVs}, \code{GetIVs}, \code{ivPtr}, \code{ivType}).


\paragraph{Gnome Objects}

The Gnome/Gtk+ Object System provides basic object-oriented programming and requires to write unsafe and unchecked code. Despite that this system cumulates all the problems of Ooc and Dynace, it is nonetheless simple, portable and implements a prototype-based object model.
\subsection{Summary}

\Cos seems to be unique by the set of features it provides to the C programming language without requiring a third party preprocessor or compiler nor any platform or compiler specific  feature. The library approach allowed to explore rapidly some object models and to select the most appropriate one fulfilling the best the aimed principles: {\em simplicity, extensibility, reusability, efficiency} and {\em portability}. Moreover, the list of features is complete and consistent: {\em augmented syntax to support object-oriented programming, uniform object model with extended metaclass hierarchy, multi-methods, fast generic delegation, design by contract, properties and key-value coding, exceptions, ownership and closures}.
\Cos features have been optimized from the design point of view, but for the sake of simplicity and portability, code tuning has never been performed 
and lets some room for future improvement. The $8000$ lines of source code of \Cos can be downloaded from sourceforge.net under the LGPL license.

\end{document}